\RequirePackage{fix-cm}
\documentclass{svjour3}                     
\smartqed  
\usepackage{graphicx}
\usepackage{color}
\usepackage{comment}
\usepackage{natbib}
\setcitestyle{authoryear}
\usepackage{amssymb}
\usepackage{amsmath}
 \journalname{Celestial Mechanics and Dynamical Astronomy}
\begin{document}

\title{Orbit determination of the moons of the Pluto-Charon system
\thanks{FK 8161, Theoretical and Computational Astrophysics, University of Patras}
}
%

\subtitle{}


\author{Dionysios Gakis         \and
        Konstantinos N. Gourgouliatos 
}


\institute{D. Gakis \at
              University of Patras \\
              Department of Physics\\
              Patras, Rio, 26504, Greece
              \email{dgakis@upnet.gr}           
           \and
           K.N. Gourgouliatos \at
              University of Patras \\
              Department of Physics\\
              Patras, Rio, 26504, Greece
              \email{kngourg@upatras.gr} 
}

\date{Received: date / Accepted: date}

\maketitle

\begin{abstract}
Four small moons (Styx, Nix, Kerberos and Hydra) are at present known to orbit around the barycenter of Pluto and Charon, which are themselves considered a binary dwarf-planet due to their relatively high mass ratio. The central, non-axisymmetric potential induces moon orbits inconvenient to be described by Keplerian osculating elements. Here, we report that observed orbital variations, may not be the result of orbital eccentricities or observational uncertainties, but may be due to forced oscillations caused by the central binary. We show, using numerical integration and analytical considerations, that the differences reported on their orbital elements, may well arise from this intrinsic behavior rather than limitations on our instruments.
\end{abstract}

\section{Introduction}
\label{intro}
Since the discovery of the dwarf planet Pluto by Clyde Tombaugh in 1930 and its moon Charon in 1978 \citep{Christy:1978}, four more moons have been detected in the system: Nix and Hydra discovered in 2005 \citep{Weaver:2006}, Kerberos in 2011 \citep{Showalter:2011} and Styx in 2012 \citep{Showalter:2012}. The flyby of New Horizons spacecraft in 2015 did not reveal any other moon orbiting the system \citep{Weaver:2016}. Within observational limits, currently only these 5 moons are believed to be present in the vicinity of Pluto, thus forming a 6-body system. Near the orbit of the outermost moon and inner to it, a possible undetected moon could only have a maximum radius of 1.7 km \citep{Weaver:2016}.

The mass ratio between Pluto and Charon is approximately 8:1, with the center of mass lying outside the volume of Pluto, which characterizes these two objects as a binary planet \citep{Stern:2015}. Because of this, the Pluto system is receiving substantial attention, since it can be a useful model for the orbits of exoplanets around binary, or even multiple, stars, or generally to provide insight into the stability of orbits around a binary system \citep{Dvorak:1986, Holman:1999}. The study of Pluto’s moon system dynamics can provide invaluable information regarding their formation and early life as well \citep{Woo:2018, Bromley:2020,Bromley:2021}. The four smaller moons are all located beyond the orbit of Charon, ordered in increasing distance from the barycenter: Styx, Nix, Kerberos and Hydra. 

Here, we study the orbits of the members of the 6-body system. Our main aim is to assess the impact of the non-axisymmetric potential generated by Pluto and Charon to the four smaller moons. We study an orbital parameter space so that we can describe the circumbinary motions of the moons. This allows us to distinguish the effects by the central binary as opposed either to a fiducial single body located at the center of mass of the system, or the approximation of two concentric rings representing Pluto and Charon that has been employed previously \citep{Showalter:2015}. Meanwhile, we re-examine the orbital uncertainties given in literature and we disentangle the orbital variability that is due to the fact that the potential arises from the Pluto-Charon binary and the intrinsic orbital eccentricity of the smaller moons. We approach these effects by a semi-analytical approach based on the work of \cite{Lee:2006} and by direct numerical integration using an n-body simulation.

 This paper is organized as follows. Section \ref{sec:2} contains an observational and Section \ref{sec:3} a theoretical summary of Pluto’s moon system. We continue with the description of the n-body integration method and the assumptions made for the simulations (Section \ref{sec:4}). We then present and analyze our results in Section \ref{sec:5}. We conclude in Section \ref{sec:6}.

\section{Observational background}
\label{sec:2}
Pluto orbits along with Charon around their center of mass nearly circularly every $6.3872273 \pm 0.0000003$ days \citep{Buie:2012}. A synchronous rotation has also been observed; Pluto and Charon are tidally locked \citep{Buie:2012}. Styx, Nix, Kerberos and Hydra have compact orbits, also nearly circular ($e\approx 10^{-3}$) and aligned with Pluto-Charon’s orbital plane ($i<1^{\circ}$). Styx, Nix and Hydra are likely in a three-body resonance, similar to Galilean moons’ resonance. This Laplace-like resonance relates the mean longitudes $\lambda_S$, $\lambda_N$, $\lambda_H$ of Styx, Nix and Hydra, respectively, as $\Phi_L=3\lambda_S-5\lambda_N+2\lambda_H \approx 180^{\circ}$, where $\Phi_L$ is the orbital phase. The ratio of the synodic periods of Nix-Hydra and Styx-Nix is then almost 3:2, but none of them appears to be in exact resonance with Charon \citep{Showalter:2015}. A second, probably inactive, resonance was proposed by \cite{Showalter:2015} too, involving Styx, Nix and Kerberos. The synodic period ratio of Nix-Kerberos and Styx-Nix is 43:42 in this case and is reminiscent of Uranus’ inner-ring moons.

The small moons have irregular shapes \citep{Porter:2016}. They are best described as triaxial ellipsoids, which is related to their chaotic rotation without a fixed rotation period and axis, \citep{Showalter:2015}. Torques by Pluto and Charon are the main culprits for those irregularities. In this study though, moons are modelled as point masses, due to their small dimensions.

Table \ref{tab:1} lists some orbital parameters for Pluto and its moons, which are in principle osculating Keplerian elements because orbits around two masses are characterized by strong perturbations, as we are going to discuss in detail in Section \ref{sec:3}. Note that the orbital axes uncertainties arising from previous theoretical works are on the order of some tens of km or less, which allow us for robust estimations of the system’s evolution. Nevertheless, we notice that reported orbital properties may differ in other works, like \cite{Brozovic:2015}. These discrepancies appear for all moons, with Styx being the most profound case at around 0.57\% deviation in estimated orbital axes.

\begin{table}
\caption{Orbital parameters for Pluto’s moon system. The semi-major axes are given with respect to the system's center of mass. The masses are based on the results of \cite{Buie:2012} and all the other  parameters are adopted from \cite{Showalter:2015}. } 
\label{tab:1}       
\begin{tabular}{llllll}
\hline\noalign{\smallskip}
Object & Mass  & Semi-major axis & Period  & Eccentricity  & Inclination \\
 & ($10^{16}$ kg) & (km)  & (days) &  ($10^{-3}$) & ($^{\circ}$) \\
\noalign{\smallskip}\hline\noalign{\smallskip}
Pluto & 1,303,000 & 2,126  & 6.3872273 & 0.000 & 0.000 \\
Charon & 158,600 & 17,470  & 6.3872273 & 0.000 & 0.000 \\
Styx & 0.06 & 42,656 $\pm$ 78 & 20.16155 $\pm$ 0.00027 & 5.787 $\pm$ 1.144 & 0.809 $\pm$ 0.162 \\
Nix & 4.5 & 48,694 $\pm$ 3 & 24.85463 $\pm$ 0.00003 & 2.036 $\pm$ 0.050 & 0.133 $\pm$ 0.008 \\
Kerberos & 0.1 & 57,783 $\pm$ 19 & 32.16756 $\pm$ 0.00014 & 3.280 $\pm$ 0.200 & 0.389 $\pm$ 0.037 \\
Hydra & 4.8 & 64,738 $\pm$ 3 & 38.20177 $\pm$ 0.00003 & 5.862 $\pm$ 0.025 & 0.242 $\pm$ 0.005 \\
\noalign{\smallskip}\hline
\end{tabular}
\end{table}

\section{Theoretical background}
\label{sec:3}

\subsection{Orbit description}
\label{sec:3.1}
Three types of orbits can be present around a binary system \citep{Dvorak:1986, Holman:1999}. One type of orbit is around the binary center of mass, so that the object is orbiting both members of the binary (P-type orbits). The second kind, called S-type, happens when the external object orbits around only one of the members of the binary, like a satellite. The third type can be an object located in the vicinity of the Lagrange points L$_4$ or L$_5$. However, the co-orbital motion corresponding to the last case is feasible only for a mass ratio between the two members of the binary being at least approximately $26:1$, which is not the case for the Pluto-Charon binary system \citep{Murray:1999}. S-type orbits in the Pluto-Charon system have already investigated in e.g. \cite{Giuliatti-Winter:2010}. Given that all four moons are in circumbinary orbits, in this paper only P-orbits will be examined.

Styx, Nix, Kerberos and Hydra orbit around the system’s barycenter, which is very close to Pluto-Charon’s barycenter, due to the low masses of the four smaller moons. However, describing their orbits using Keplerian osculating elements (i.e. like orbits around a central mass) is not convenient because of the time-dependent gravitational potential of the central binary, the relatively high central mass ratio and the proximity of moons to Pluto and Charon. Thus, an elliptic orbit described by Kepler’s laws provides only a first approximation on long timescales  by taking time-averages and is not sufficiently accurate for short-term analysis of the system, described in this paper \citep{Lee:2006, Youdin:2012}.

Here we present the basic elements of the epicyclic theory for circumbinary orbits as it was developed by \cite{Lee:2006}, Lee-Peale theory hereafter, and was further discussed by other researchers, e.g.~\cite{Woo:2020, Bromley-Kenyon:2020}, so that we can apply it to the Pluto-Charon and moons system. Lee-Peale theory is suitable for separating the effects of the two-body potential, and that is why we adopt it here. It holds for nearly coplanar orbits ($i\ll \pi /2$) and zero eccentricities for the central objects. Moons are treated as test particles of negligible mass.

 We first define the guiding center, which is a reference point in circular orbit about the barycenter. Coordinates are determined to account for deviations from this uniform motion. The particle's orbit around two central objects is described by the superposition of the guiding center's circular orbit (radius $R_S$), an epicyclic motion, forced oscillations by the non-axisymmetric potential and vertical motion. 
 
 Let us assume that Pluto and Charon orbits lie on the $z=0$ plane, and their center of mass is set to $(0, 0, 0)$. Their separation is $a_{PC}$ and their circular frequency, or mean motion, is $n_{PC}=[G(m_{P}+m_{C})/a_{PC}^3]^{1/2}$. Thus, adopting cylindrical coordinates, their respective positions are $\mathbf{r_P}=(a_P, \phi _C+\pi, 0)$ and $\mathbf{r_C}=(a_C, \phi _C, 0)$, where $a_P=a_{PC}m_{C}/(m_{P}+m_{C})$, $a_C=a_{PC}m_{P}/(m_{P}+m_{C})$ and $\phi _C(t)=n_{PC}t+\phi '$ ($\phi '$  a constant). The combined gravitational potential by Pluto and Charon at a point $\mathbf{r}=(R, \phi, z)$ from the system’s barycenter can be expressed as:
\begin{eqnarray}  \Phi(\mathbf{r})= -\Big{\{}\frac{Gm_P}{\left[z^2+R^2+a_P^2+2Ra_P\cos (\phi -\phi_C(t))\right]^{1/2}}\nonumber \\
+ \frac{Gm_C}{\left[z^2+R^2+a_C^2-2Ra_C\cos (\phi -\phi_C(t))\right]^{1/2}}\Big{\}}
\end{eqnarray}
\begin{equation} =\sum_{k=0}^\infty \left [\Phi_{0k}(R)-\frac{1}{2}\left(\frac{z}{R}\right)^2\Phi_{2k}(R)+...\right] \cos k\left(\phi-\phi_c\right)\end{equation}
where
\begin{eqnarray} \Phi_{jk}(R)&=&-\frac{2-\delta _{k0}}{2}\left[\frac{m_{C}}{(m_{P}+m_{C})}b_{(j+1)/2}^k(\alpha_C)+(-1)^k\frac{m_{P}}{(m_{P}+m_{C})}b_{(j+1)/2}^k(\alpha_P)\right]\nonumber \\ &\times& \frac{G(m_{P}+m_{C})}{R}\end{eqnarray}
$\alpha_C=a_C/R$ and $\alpha_P=a_P/R$ are the normalized distances, $\delta _{k0}$ is Kronecker's delta and $b_{s}^k$ are the Laplace coefficients. 

This non-axisymmetric potential is the reason why only the orbits of Pluto and Charon can be accurately described by Kepler’s laws. The gravitational field of a binary planet (or star) causes orbital behavior similar to the one arising from a single non-spherical (oblate) body with a large J$_2$ term \citep{Murray:1999}. 

The resulting solution is the linear combination of harmonic oscillator modes ($\kappa, \lambda$ being constant phase angles):
\begin{equation} R(t)=R_S\left[1-e_{free}\cos \left(v_et+\kappa\right)+\sum_{k=1}^\infty C_k \cos \left[k\left(n_{PC}-n_{S}\right)t\right]\right] \end{equation}
\begin{equation} \phi (t)=n_{S}\left[t+\frac{2e_{free}}{v_e}\sin \left(v_et+\kappa\right)+\sum_{k=1}^\infty  \frac{D_k}{k\left(n_{PC}-n_{S}\right)} \sin \left[k\left(n_{PC}-n_{S}\right)t\right]\right] \end{equation}
\begin{equation} z(t)=iR_S\cos \left[v_it+\lambda\right] \end{equation}

The guiding center's mean motion $n_{S}$, the epicyclic frequency $v_e$ and the vertical frequency $v_i$ are defined as follows:
\begin{equation}n_{S}^2=\frac{1}{R_S}\frac{d\Phi _{00}}{dR}\Bigg|_{R_S}\end{equation}
\begin{equation}v_{e}^2=R_S\frac{dn _{S}^2}{dR}\Bigg|_{R_S}+4n _{S}^2\end{equation}
\begin{equation}v_{i}^2=\frac{1}{z}\frac{d\Phi }{dz}\Bigg|_{z=0,   R_S}\end{equation}
Factors $C_k$ and $D_k$ represent the amplitudes of the oscillations:
\begin{equation}
C_k=\left[\frac{1}{R_S}\frac{d\Phi _{k}}{dR}\Bigg|_{R_S}-\frac{2n_S\Phi_k}{R_S^2\left(n_{PC}-n_{S}\right)}\right]\frac{1}{v_e^2-k^2\left(n_{PC}-n_{S}\right)^2}
\end{equation}
\begin{equation}
D_k=2C_k+\frac{\Phi_k}{R_S^2n_S \left(n_{PC}-n_{S}\right)}
\end{equation}
Free eccentricity $e_{free}$ is an independent parameter in the solution and stands for the amplitude of the epicyclic motion at a rate determined by $v_e$. Accordingly, $i$ outlines the vertical oscillations at frequency $v_i$. Forced oscillations caused by the rotating non-axisymmetric components of the central binary extend to values set by the $C_k$ term and occur at frequencies  $|k(n_S-n_{PC})|$. Periapsis and ascending nodes are precessing with nearly identical rates but in opposite directions. Periapsis precession rate is $n_S-v_e$ (prograde) and nodal precession rate is $n_S-v_i$ (retrograde).

The above solution is valid for low free eccentricities and inclinations for particles around a circular binary system. Errors increase by $i^2$ and at the second order of $e_{free}$. \cite{Leung:2013} extended the Lee-Peale theory for eccentric central orbits and found additional oscillation terms. Since plutocentric Charon's orbital eccentricity is at present consistent with zero \citep{Buie:2012}, we do not examine this theory here. There are other theoretical approaches to this problem as well, e.g.~using Hamiltonian mechanics in \cite{Georgakarakos:2015} and in \cite{Woo:2020}, but we choose not to analyze them as they result in solutions compatible with Lee-Peale theory.

\subsection{Stability limits}
\label{sec:3.2}

Pluto-Charon’s Hill sphere establishes the region of their gravitational dominance over the Sun. Its radius is given by:
\begin{equation} R_{H,PC}=\left(\frac{m_{P}+m_{C}}{3M_{\odot}}\right)^{1/3}r_{P}(1-e_{P})
\end{equation}
where $r_P$ is the Pluto-Charon's distance from the Sun and $e_P$ the eccentricity of Pluto’s orbit around the Sun. Since the moons of Pluto orbit at distances $~0.007-0.011R_{H,PC}$, all of them are located well inside the system’s Hill sphere of influence. As a result, the gravity of the Sun is often neglected when studying the stability of the system, being present sometimes only as a perturbation \citep{Michaely:2017}.

Another solar-tidal stability limit is also defined for moons in near-circular orbits (Szebehely criterion). The critical distance $R_{Sz}$ for this criterion is one third of the Hill radius. Calculations suggest that at distances $R_{Sz}< a < R_H$ instability is possible, whereas for $a >R_H$ instability always appears \citep{Szebehely:1967,Szebehely:1978}. For that reason, Szebehely criterion is often adopted instead of Hill criterion, as it can restrict further our search for possible moons, e.g. \cite{Stern:1991}, \cite{Steffl:2006}. 

We may also define a moon’s Hill radius, inside of which every moving object is bound to the moon (beyond of it an object is bound to the central binary). For a $j$ moon, it is given, similarly to previously, by:
\begin{equation} R_{H,j}=\left(\frac{m_{j}}{3(m_{P}+m_{C})}\right)^{1/3}a_{j} 
\end{equation}

The stability of a moon’s orbit is not only affected by the central binary; other moons may have an effect, as well. This is why we specify the mutual Hill radius for $j$ and $q$ moons:
\begin{equation} R_{H,jq}=\left(\frac{m_{j}+m_{q}}{3(m_{P}+m_{C})}\right)^{1/3}\overline{a} \,,
\end{equation}
where $\overline{a}$ is found by taking the average of the semi-major axes. For an easier comparison, we use the normalized separation, $K=|a_j-a_q|/R_{H,jq}$. Theoretical analyses suggest that for a system consisting of many bodies (like Pluto’s system), $K\approx10-12$ is required to ensure stability at a timescale comparable to the lifetime of the Sun \citep{Kenyon:2019}. In the case of Pluto’s moon system, although the moons generally do not exceed these limits, one pair (Kerberos-Hydra) is marginally stable, at $K=11$, and other two (Styx-Nix and Nix-Kerberos) are slightly above them.

As for the innermost coplanar prograde stable orbit around Pluto and Charon, previous considerations suggest that it belongs in the range $1.7-2.0a_{PC}$. Particles placed inside these regions will be ejected from the system in less than about $10^3$ years. Styx is just beyond the innermost limit, at a distance of $2.2a_{PC}$. For a polar orbit, the minimum semi-major axis for a stable orbit around the central binary is somewhat larger, $2.2a_{PC}$, whereas it reduces to $1.3a_{PC}$ for a retrograde orbit \citep{Doolin:2011}.

\subsection{Effects by the Sun}
\label{sec:3.3}

As mentioned previously, the gravitational effect of the Sun is only present as a perturbation \citep{Michaely:2017}. In the Sun-Pluto-Charon reference system, Pluto and Charon orbit around their center of mass in circular orbits (inclination is $119^{\circ}$ with respect to Pluto’s orbital plane around Sun) and their center of mass orbits the Sun. The outer body (Sun) causes perturbations in the binary by causing quasi-periodic oscillations \citep{Michaely:2017}. These oscillations, called Kozai-Lidov (KL) cycles, disturb the outer regions of the binary system by changing orbital inclinations and eccentricities, but conserving the semi-major axes. On the other hand, nodal oscillations caused by the binary dominate over Kozai-Lidov (KL) oscillations at smaller distances (nodal periods are shorter than KL periods). Beyond a critical distance $a_{crit}$ though, the orbit of a moon is irregular (high eccentricity and inclination). $a_{crit}$ has been calculated in \cite{Michaely:2017} by equating KL rate with inner binary precession rate, and the result yields ($e_m$ being the moon's orbital eccentricity):
\begin{equation} a_{crit}=\sqrt[5]{\frac{3}{8}\frac{a_C^2a_P^3m_P m_C (1-e_P^2)^{3/2}(5\cos^2i-1)
}{(m_P+m_C)(1-e_S^2)^2M_{\odot}}}\,. \end{equation}
This is the distance where the inner binary precession timescales are equal to KL precession timescale. Inner to this distance, only binary precession is important, while beyond it, binary precession is neglected. The resulting value of $a_{crit}$ for Pluto's system is $5.24\times 10^8$ m, which is smaller than Hill radius but larger than Hydra’s orbital axis by approximately one order of magnitude. Thus, while the Sun can affect the dynamical system at some distances from the barycenter, we choose to ignore its gravitational influence, since these distances exceed by far the outermost moon’s semi-major axis. We also neglect the gravitational effect of other planets.

The Sun may also disturb some parts of Pluto’s system by its radiation pressure. Smaller particles (e.g. dust particles, produced by collisions of interplanetary debris with massive bodies) are affected by solar radiation even at distances from the Sun as far as Pluto’s. Solar radiation force is analyzed in two components: solar radiation pressure, which changes eccentricity and Poynting-Robertson drag, which causes orbital decay \citep{Burns:1979}. However, we do not account solar radiation pressure here, since the calculated timescales for the radiation pressure influence are comparable with gravitational timescales only for particles with sizes of $r\leq 1$ mm \citep{Santos:2011}. Other minor effects and non-conservative forces like planetary shadow and Yarkovsky effect \citep{Hamilton:1996} are neglected as well, given the minimal impact they have on the orbital evolution.

\section{Method}
\label{sec:4}

\subsection{N-body code}
\label{sec:4.1}

We explore orbits in Pluto’s system performing numerical calculations with a gravitational n-body integrator, in a Python 3.9.6 IDLE environment. The code, developed by Philip Mocz, utilizes the leapfrog scheme for the calculations. This ‘kick-drift-kick’ technique is widely used to solve differential equations of the form $\ddot x=f(x,t)$ present in classical mechanics (e.g. \cite{Springel:2021}). The, code evaluates pairwise the distances between all the pairs of objects of the system, and the resulting accelerations $\bf{a}$ for body $\mu$ summing over all other bodies in the system:
\begin{eqnarray}
\bf{a}_{\mu}=\sum_{\nu\neq \mu} G m_{\nu}\frac{{\bf r}_{\nu}-{\bf r}_{\mu}}{|{\bf r}_{\nu}-{\bf r}_{\mu}|^3}\,,
\end{eqnarray}
where ${\bf r}_{\nu}$ is the position vector for the $\nu$-th body. Then the velocities are evaluated for half time step $\Delta t/2$:
\begin{eqnarray}
{\bf v}_{\mu}\left(t+\frac{\Delta t}{2}\right)={\bf v}_{\mu}(t)+ {\bf a}_{\mu}(t) \frac{\Delta t}{2}\,.
\end{eqnarray}
It is followed by a full time step for the position using the newly calculated velocities:
\begin{eqnarray}
{\bf r}_{\mu}\left(t+\Delta t\right)={\bf r}_{\mu}(t)+ {\bf v}_{\mu}\left(t+\frac{\Delta t}{2} \right) \Delta t\,.
\end{eqnarray}
This is then concluded by another half time step for the velocity, where the acceleration has now been evaluated on the positions at ${\bf r}_{\nu}\left(t+\Delta t\right)$.
\begin{eqnarray}
{\bf v}_{\mu}(t+\Delta t)={\bf v}_{\mu}\left(t+\frac{\Delta t}{2}\right)+ {\bf a}_{\mu}\left(t+\Delta t \right) \frac{\Delta t}{2}\,.
\end{eqnarray}

The numerical scheme is a symplectic second-order integrator, thus offering the advantage of preserving the total energy of the system and being time-reversible.

Performing simulations for timespans comparable to the Solar System’s age is an ideal, but impractical option. Besides, our objective is to examine circumbinary motions and not to carry out a long-time analysis. For the above reasons, we mainly set the total time of the numerical calculations to approximately a few thousand years. We acknowledge that this timescale is not enough for definite conclusions about the long-term stability of Pluto’s system, but it is indeed suitable to provide a thorough description of orbits. Also, it enables us to check our input values and detect any deviations on the consistency between the results (positions and velocities) employed in other simulations, or deduced by observations. We further explore for orbital patterns on much shorter time intervals (less than a year).

\begin{table}
\caption{Case description}
\label{tab:2}       
\begin{tabular}{llll}
\hline\noalign{\smallskip}
Case & Simulated bodies & Initial velocity & Initial data source  \\
\noalign{\smallskip}\hline\noalign{\smallskip}
A & all & measured vectors & \cite{Brozovic:2015}  \\
B & all & $\upsilon = 2 \pi a/P$ & \cite{Brozovic:2015}  \\
C & all & $\upsilon = (GM/a)^{1/2}$ & \cite{Brozovic:2015}  \\
D & all & $\upsilon = (2 \pi a)/P$ & \cite{Showalter:2015}  \\
E & all & $\upsilon = (GM/a)^{1/2}$ & \cite{Showalter:2015}  \\
F & all, except combined Pluto and Charon & $\upsilon = (GM/a)^{1/2}$ & \cite{Showalter:2015}  \\
\noalign{\smallskip}\hline
\end{tabular}

\end{table}

\begin{figure*}
\begin{tabular}{p{0.49\textwidth} p{0.5\textwidth}}
  a\includegraphics[width=0.49\textwidth]{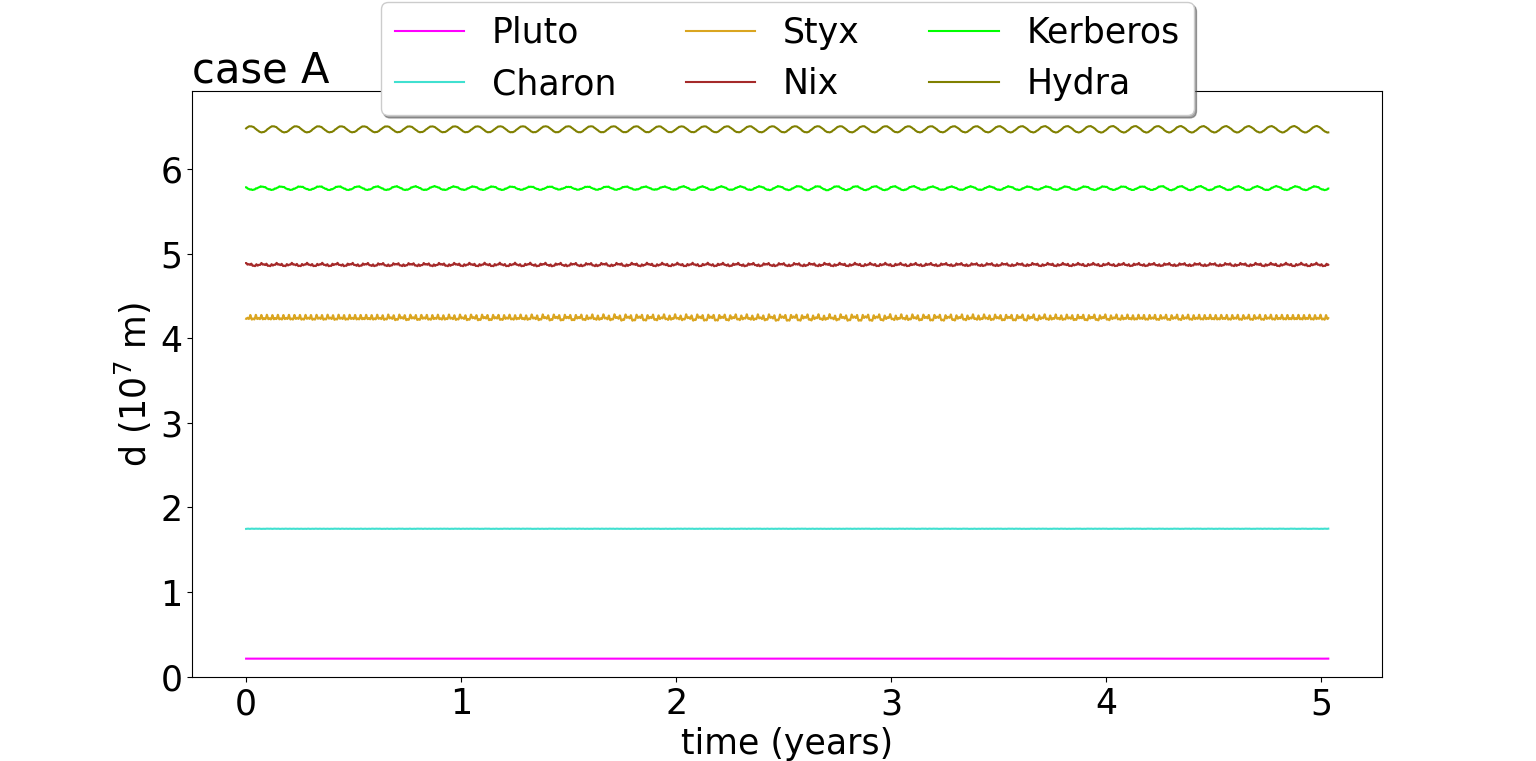}&
    b\includegraphics[width=0.49\textwidth]{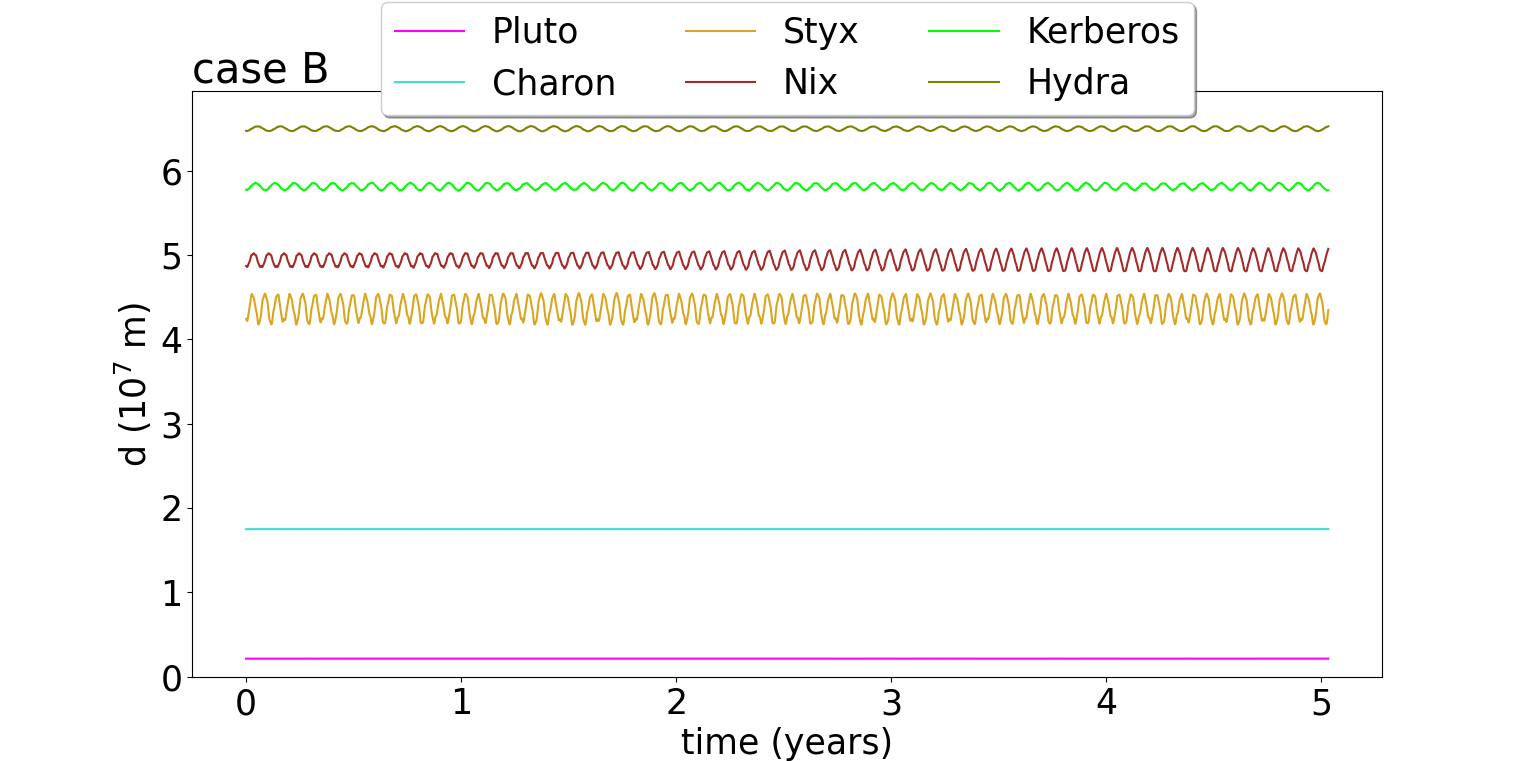} \\
     c\includegraphics[width=0.49\textwidth]{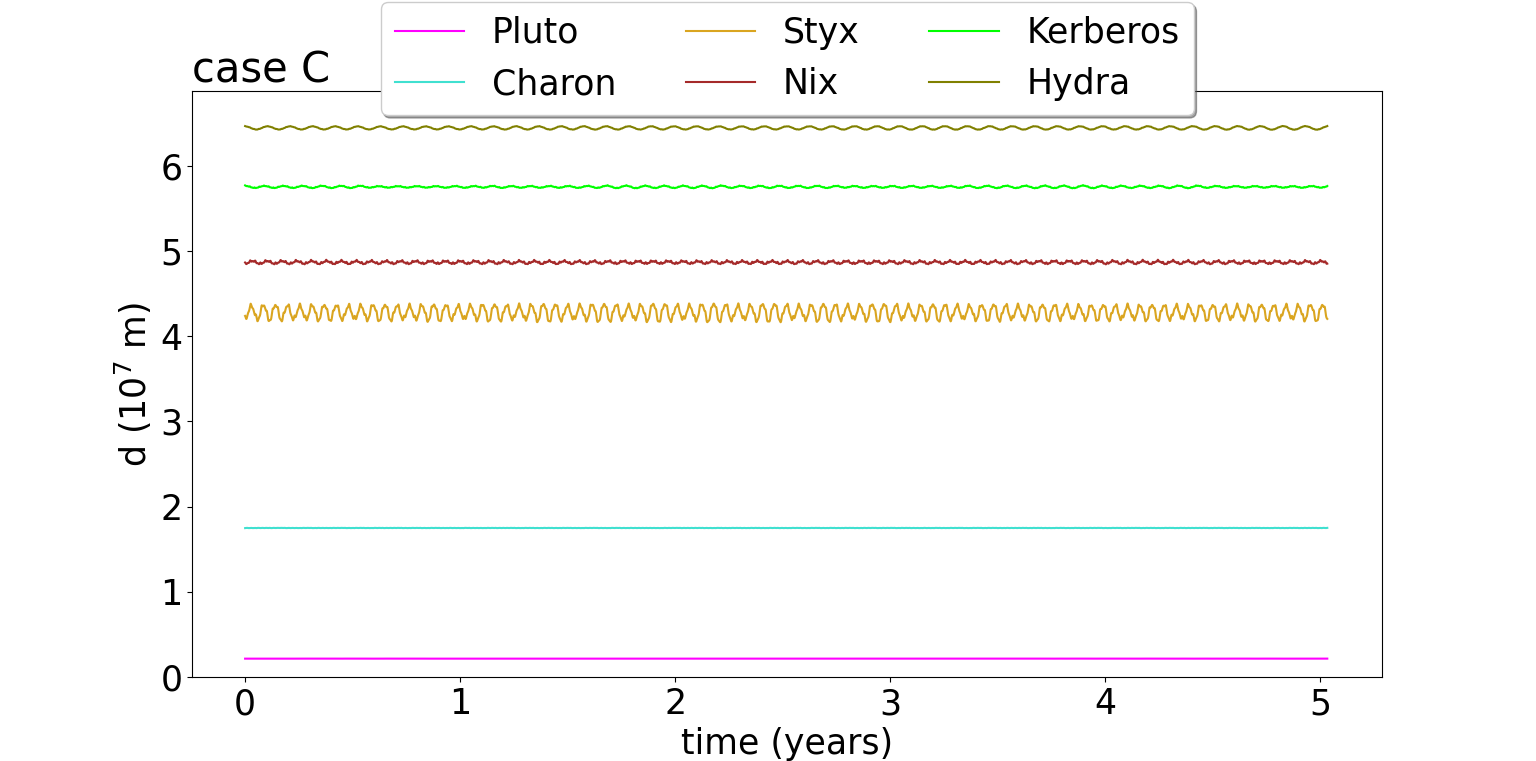} &
       d\includegraphics[width=0.49\textwidth]{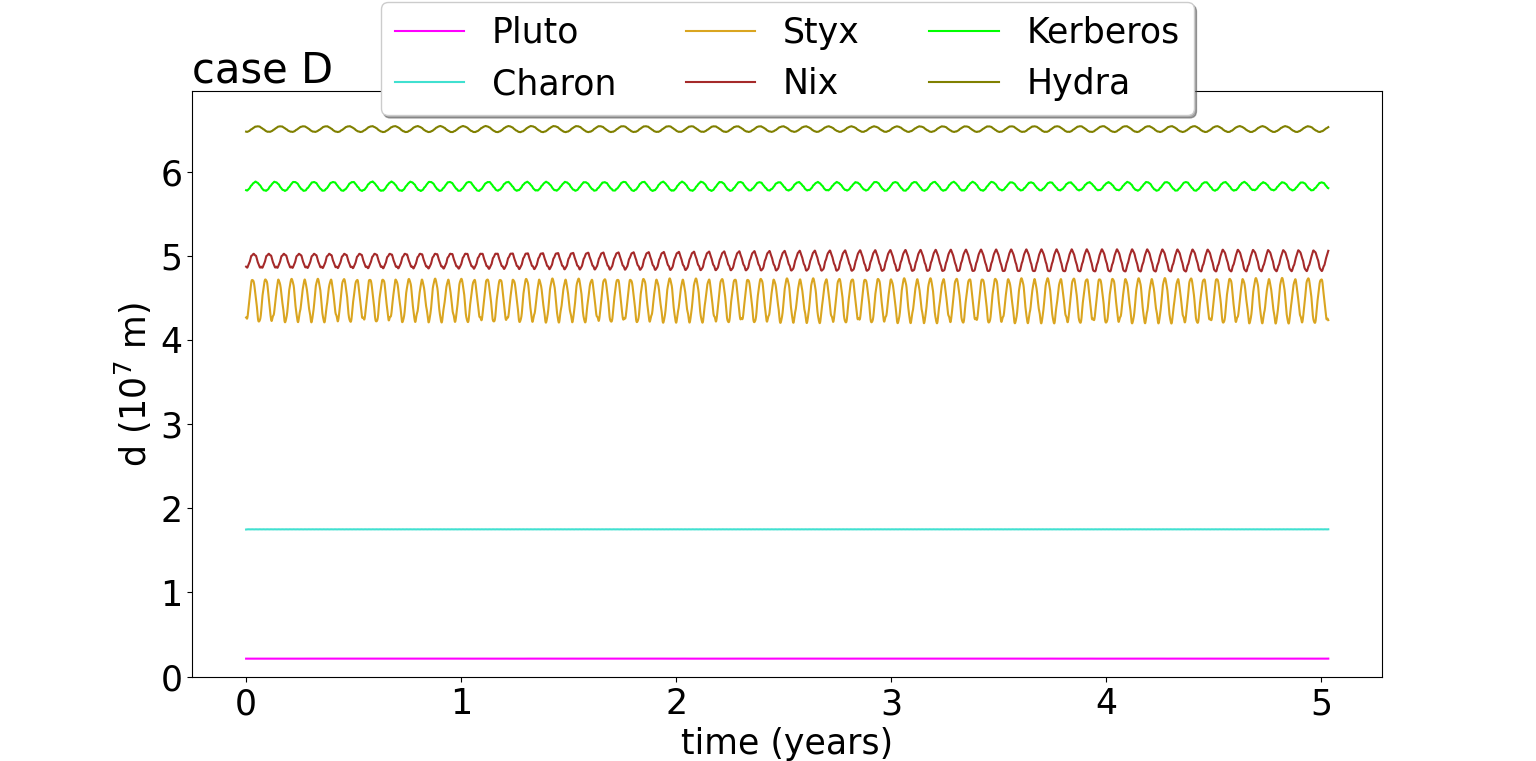} \\
    e\includegraphics[width=0.49\textwidth]{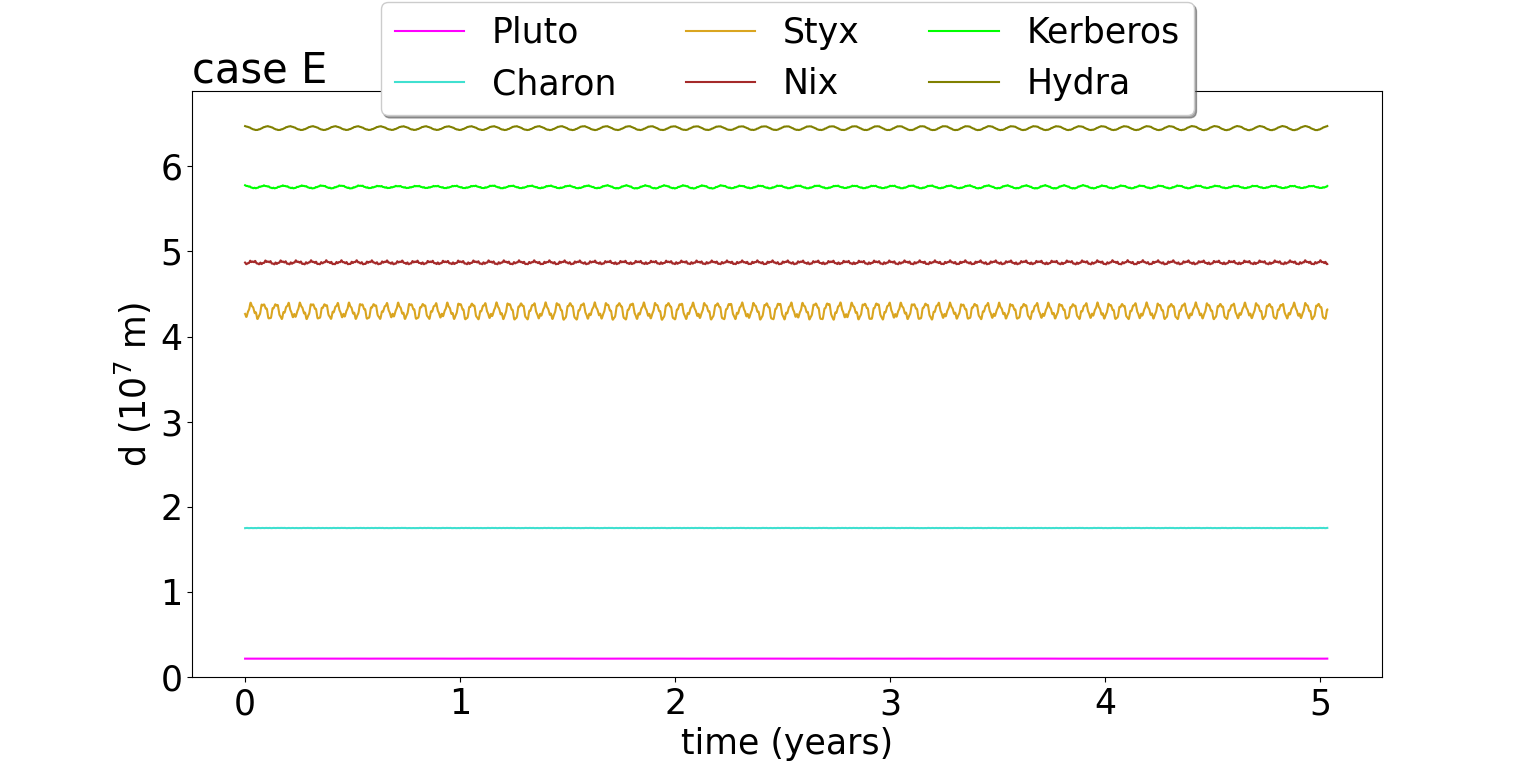} &
     f\includegraphics[width=0.49\textwidth]{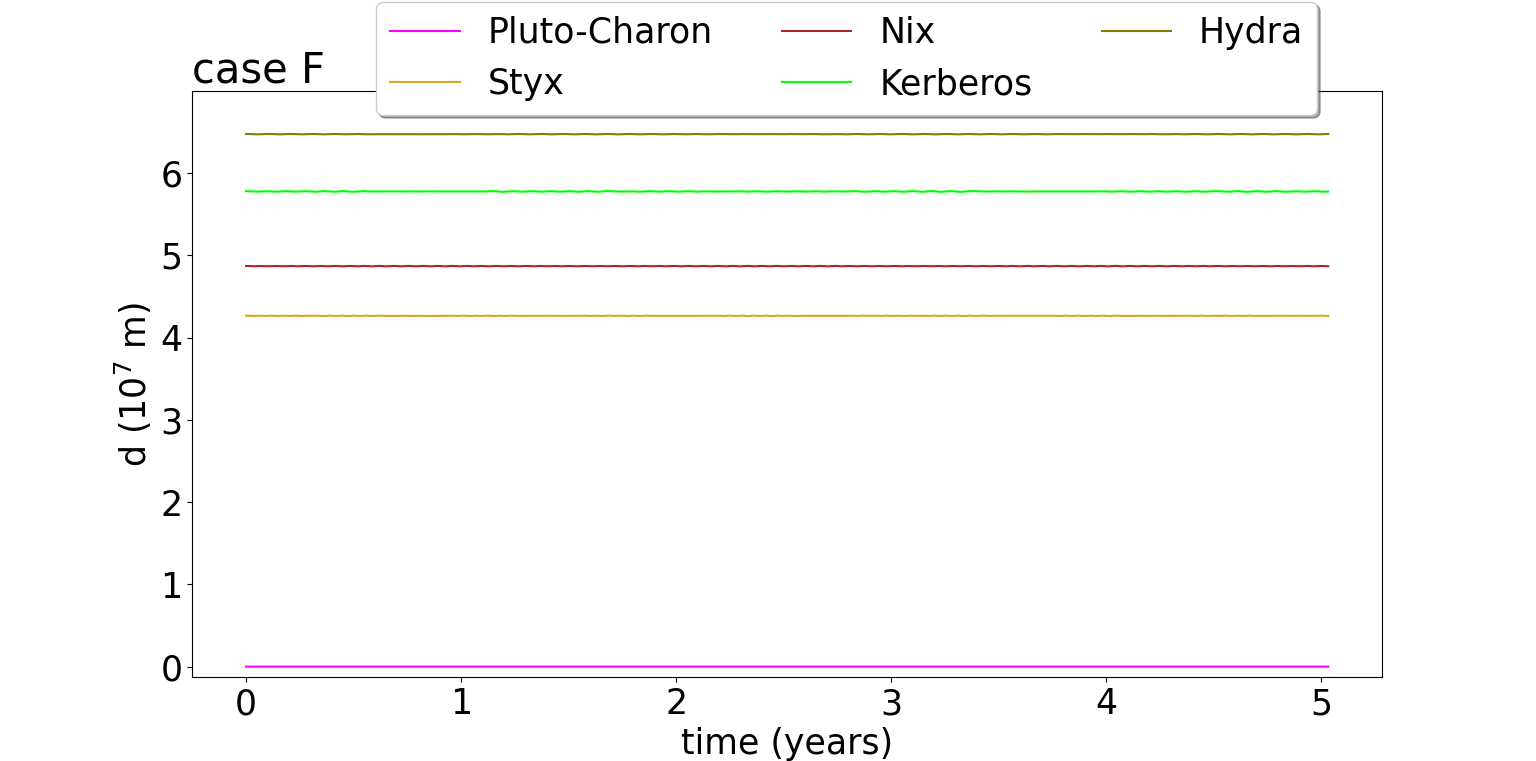}
     \end{tabular}
\caption{Variation of distance from barycenter ($d$) through a timespan of 5 years for all bodies. The cases A, B, C, D, E and F are shown in the respective panels a, b, c, d, e and f.}
\label{fig:1}       
\end{figure*}

For our simulations, the adopted timestep for calculations is given by $\Delta t=T_{PC}/N$, $T_{PC}$ being the Pluto-Charon orbital period and N an integer. Obviously, a smaller value for N results in a larger timestep, affecting the accuracy of the numerical calculations. Nevertheless, the total processing time is proportional to N. As a result, choosing an extremely large N is impractical, due to limited computational power. In fact, $N\approx100$ is considered a reasonable choice, both keeping the uncertainties within an acceptable level and maintaining the computational time relatively short ($\sim$ some hours, depending on total timespan). $N \lessapprox 50$ is thought to be a poor approximation, keeping uncertainties up to more than $0.5\%$ in some cases. Note that the errors $\Delta t$ produces depend strongly on the total time simulated. We then adopt $N=110$ for our calculations hereafter ($\Delta t=5,000$ s), unless otherwise stated. We have also performed some test simulations with $N\approx200$, though, without noticing any significant variations of the solutions (less than $0.075\%$ maximum difference).

\subsection{Initial conditions}
\label{sec:4.2}

We use \cite{Brozovic:2015} and \cite{Showalter:2015} as benchmarks for our study. In both of these studies, Keplerian elements for Pluto-Charon system were derived by best-fit orbits using observational data. This does not produce accurate results for circumbinary orbits, as we demonstrated in Section \ref{sec:3.1}, but is valuable for average estimates (it produces exact mean values). We examine how consistent are these values with the Lee-Peale theory. For that purpose we performed a series of n-body calculations, changing every time the initial conditions. 

All of our n-body integrations are limited to the 6 bodies constituting Pluto-Charon's dynamical system because the effects of the Sun or other bodies in the Solar System are insignificant for our analysis, as indicated in Section \ref{sec:3.3}. In addition, each object is modelled as point mass despite any irregularities in its shape, as our objective is to study the impact of the non-axisymmetric potential caused by the central binary. Likewise, Lee-Peale theory approximates moons considered as point-like particles.

\begin{figure*}
\begin{tabular}{p{0.49\textwidth} p{0.5\textwidth}}
  a\includegraphics[width=0.49\textwidth]{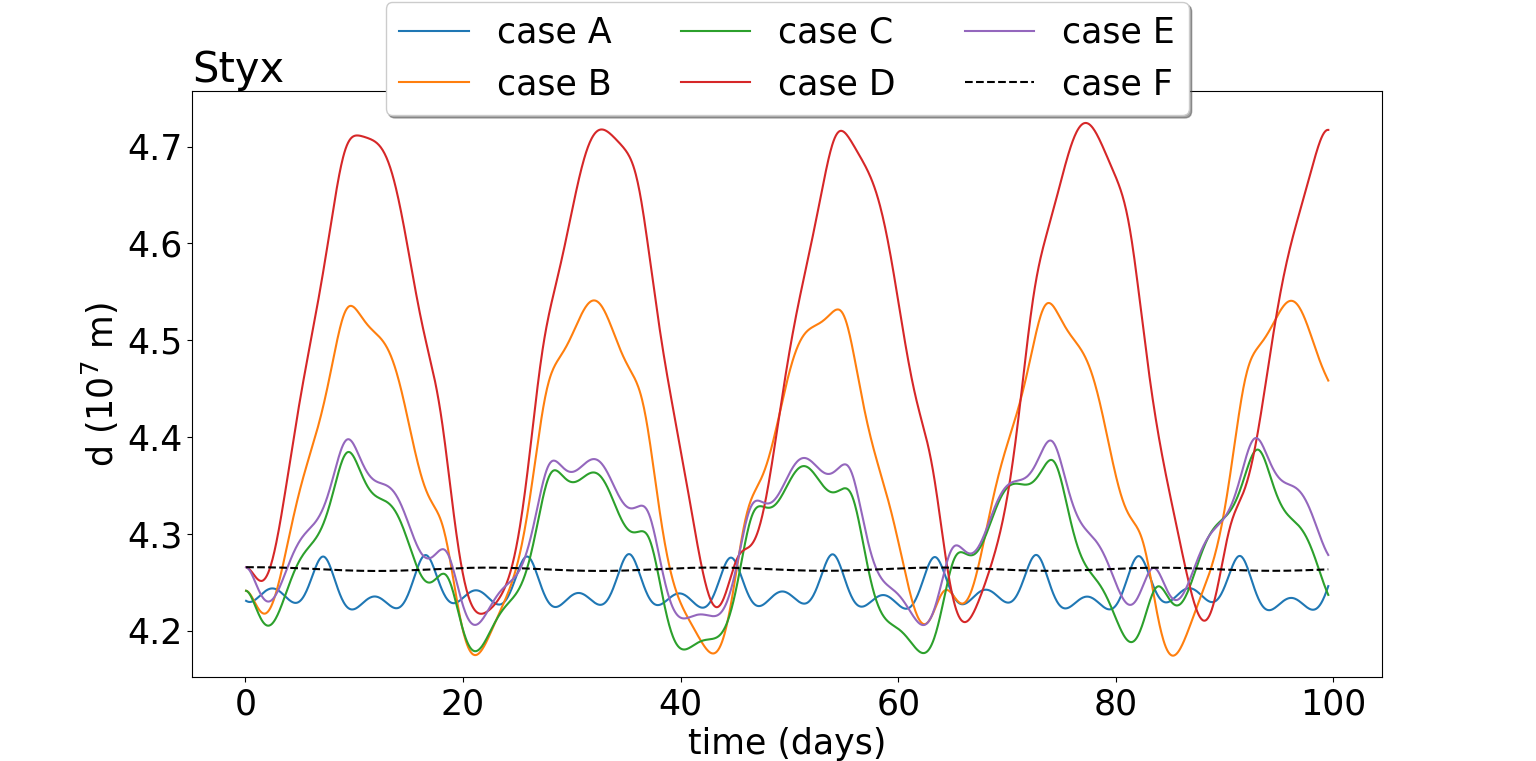}& 
    b\includegraphics[width=0.49\textwidth]{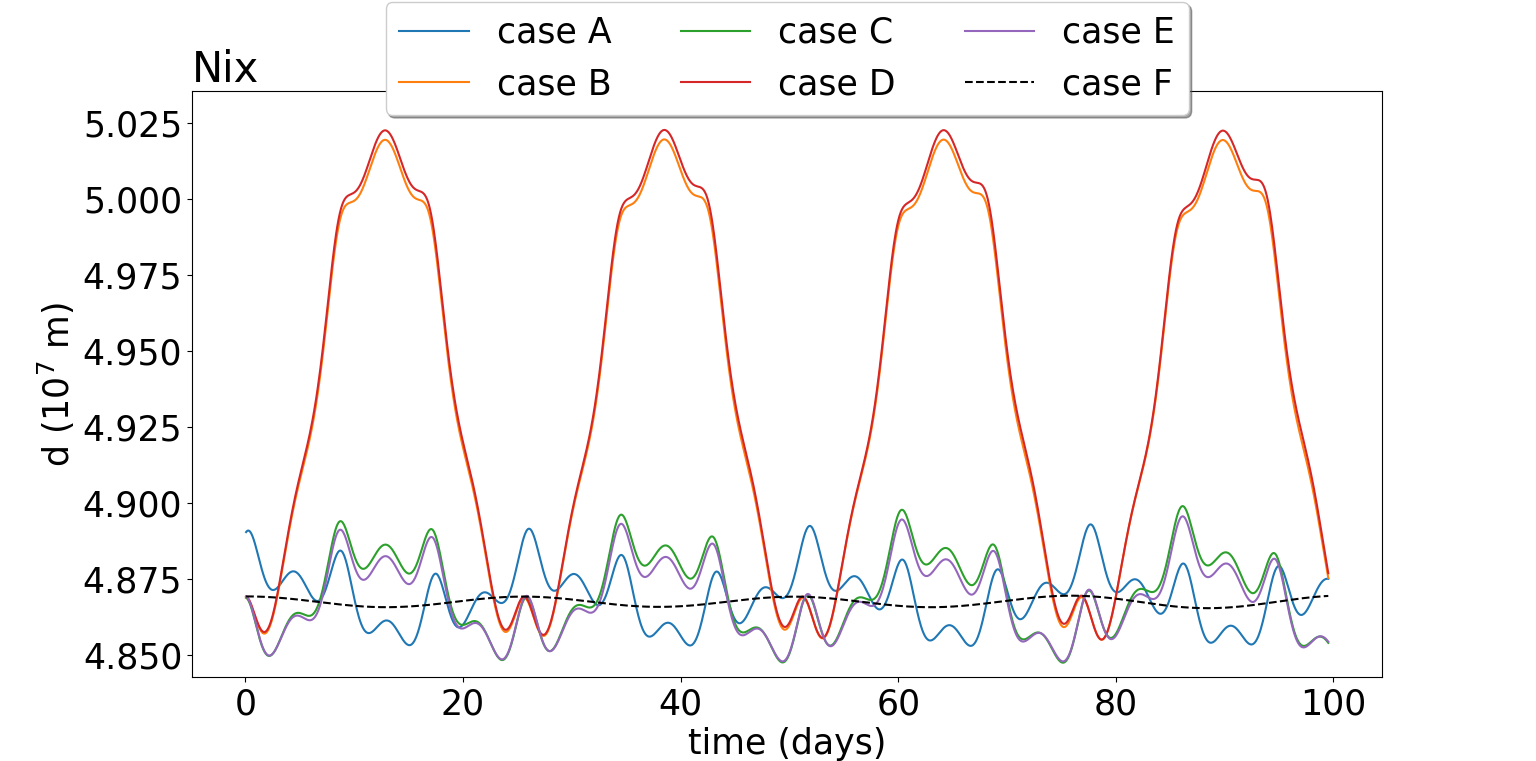} \\
     c\includegraphics[width=0.49\textwidth]{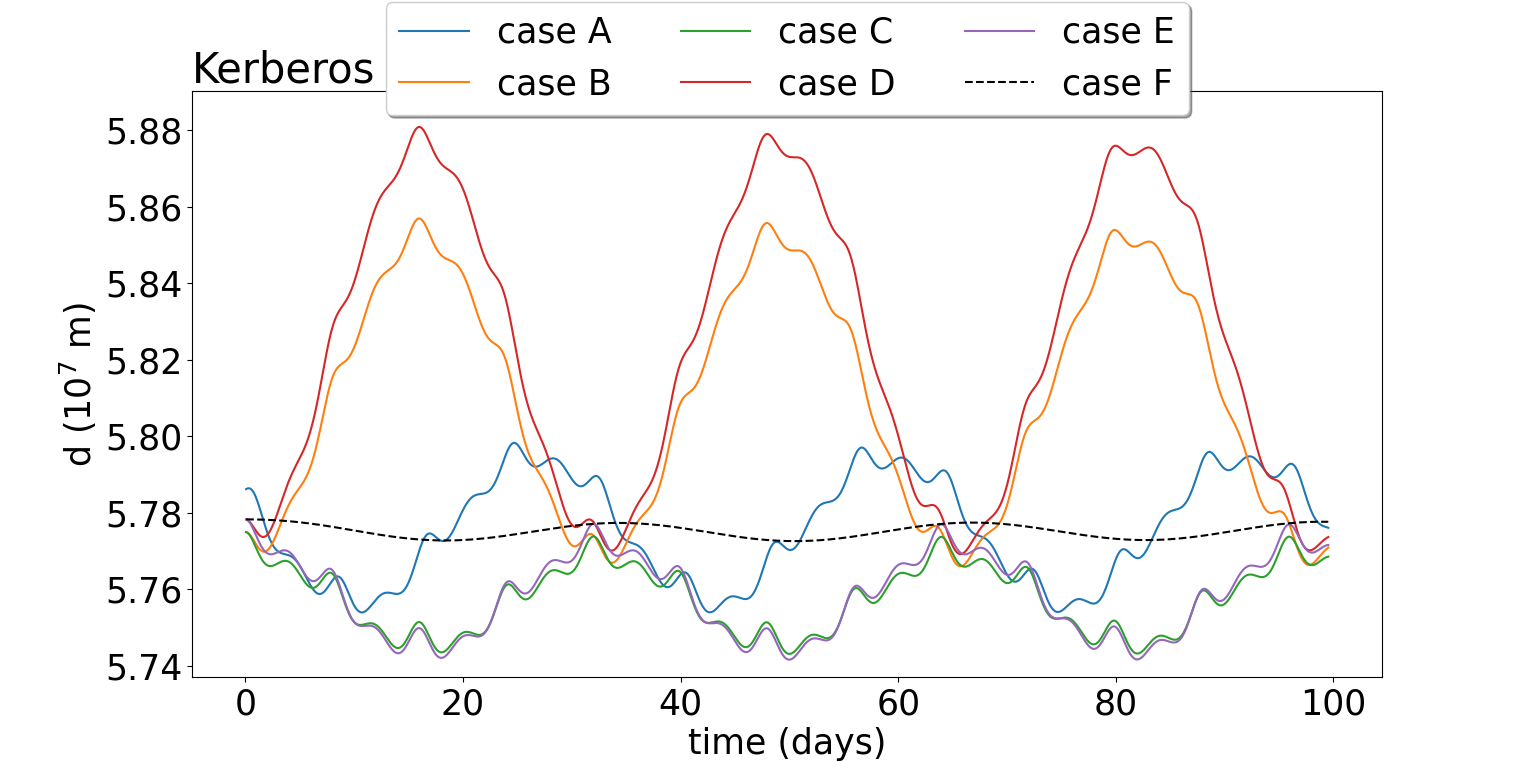} &
    d\includegraphics[width=0.49\textwidth]{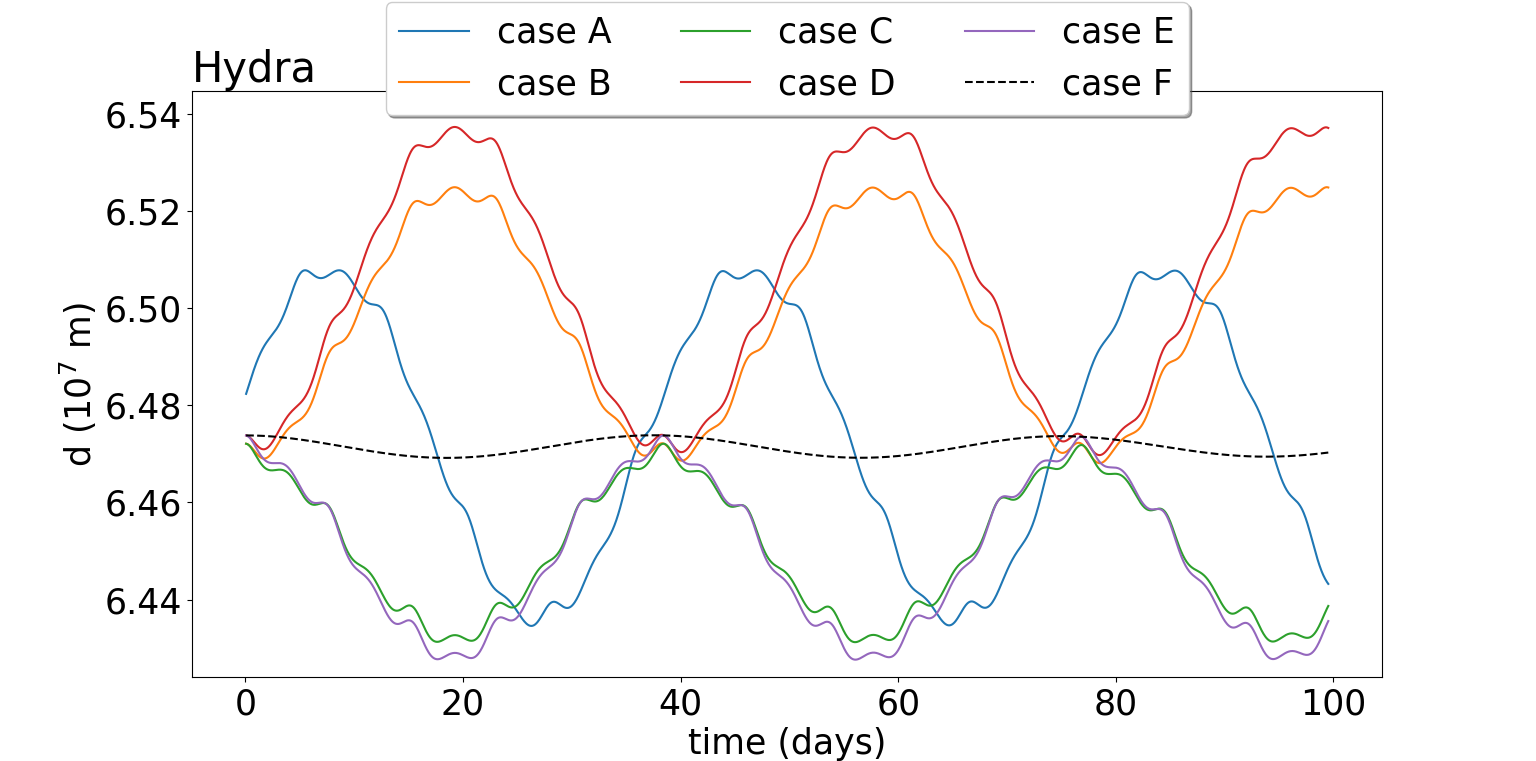} 
        \end{tabular} 
\caption{Variation of distance from barycenter ($d$) through a timespan of 100 days for Styx, Nix, Kerberos and Hydra at respective panels a, b, c and d. The cases A, B, C, D, E and F are shown.}
\label{fig:2}       
\end{figure*}

We perform six simulations, A-F, which are presented in Table \ref{tab:2}. The masses of the members of the system are the ones appearing in Table \ref{tab:1}. Our solution strategy is the following. In case A, we integrate the state vectors, as provided by \cite{Brozovic:2015} forward in time (Table 8 therein). Positions and velocities are set in the International Celestial Reference Frame relative to the system's barycenter. The epoch is January 1st 2013 TDB (Barycentric Dynamical Time). Uncertainty is at least limited to eleven decimal digits. This simulation does not restrict the orbits on a single plane, but it includes the appropriate inclination. 

In cases B and C, velocities are obtained using $\upsilon = 2\pi a/P$ and $\upsilon = (GM/a)^{1/2}$ formulae, respectively, utilizing data from \cite{Brozovic:2015} (Table 9 therein), where $M$ is the sum of the masses of Pluto and Charon. Next, in cases D and E, we follow the same procedure, now using semi-major axes and periods from \cite{Showalter:2015}. Throughout simulations B-F, the Pluto-Charon orbital plane is rotated so that it is aligned with $z=0$ plane. All other moons are set to orbit on the same plane, thus ignoring their small inclinations. Our aim in these sets of simulations is to compare the theoretical with the observational outcomes, and assess the impact of measurement errors for the latter. 

Finally, in case F we assume a point object whose mass is equal to the sum of the masses of Pluto and Charon, that is placed at the center of mass of Pluto and Charon. Essentially, we merge the binary planet into a single body. We note that there is small deviation between the center of mass of Pluto and Charon and the system's barycenter. All orbits are expressed with respect to  the system's barycenter. Regarding the four moons, we use the semi-major axes given in \cite{Showalter:2015} and velocities obtained by $\upsilon = (GM/a)^{1/2}$. This allows us to quantify and disentangle the effect of the time-dependent and non-axiymmetric potential due to Pluto and Charon to the orbits from intrinsic eccentricities that may be present in the reported orbital elements in \cite{Showalter:2015, Brozovic:2015}. We note that the combinations of periods and semi-major axes should correspond to minimal eccentricities $(e \lesssim 0.005)$.

\begin{figure*}
\begin{tabular}{p{0.49\textwidth} p{0.5\textwidth}}
  a\includegraphics[width=0.49\textwidth]{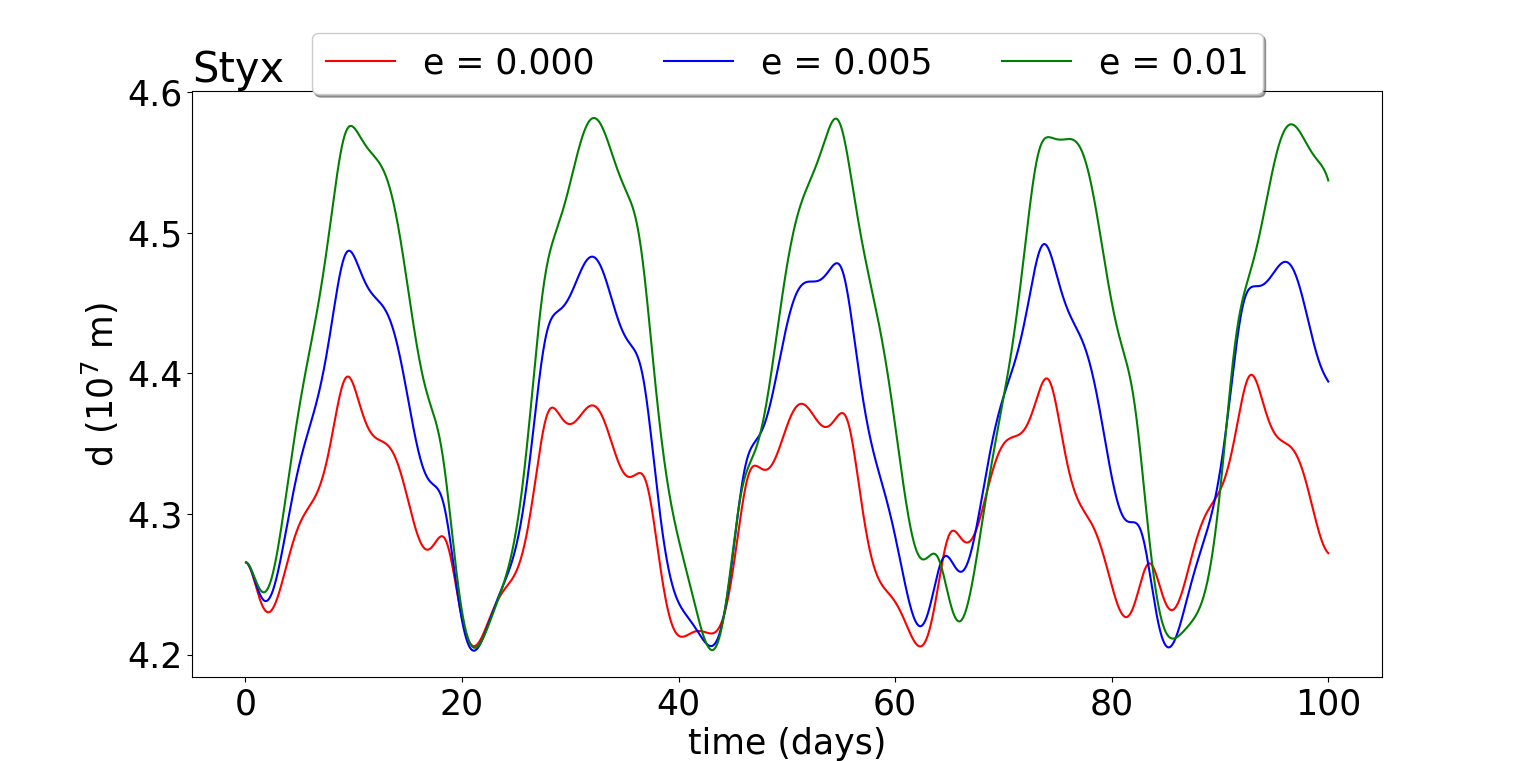} & 
    b\includegraphics[width=0.49\textwidth]{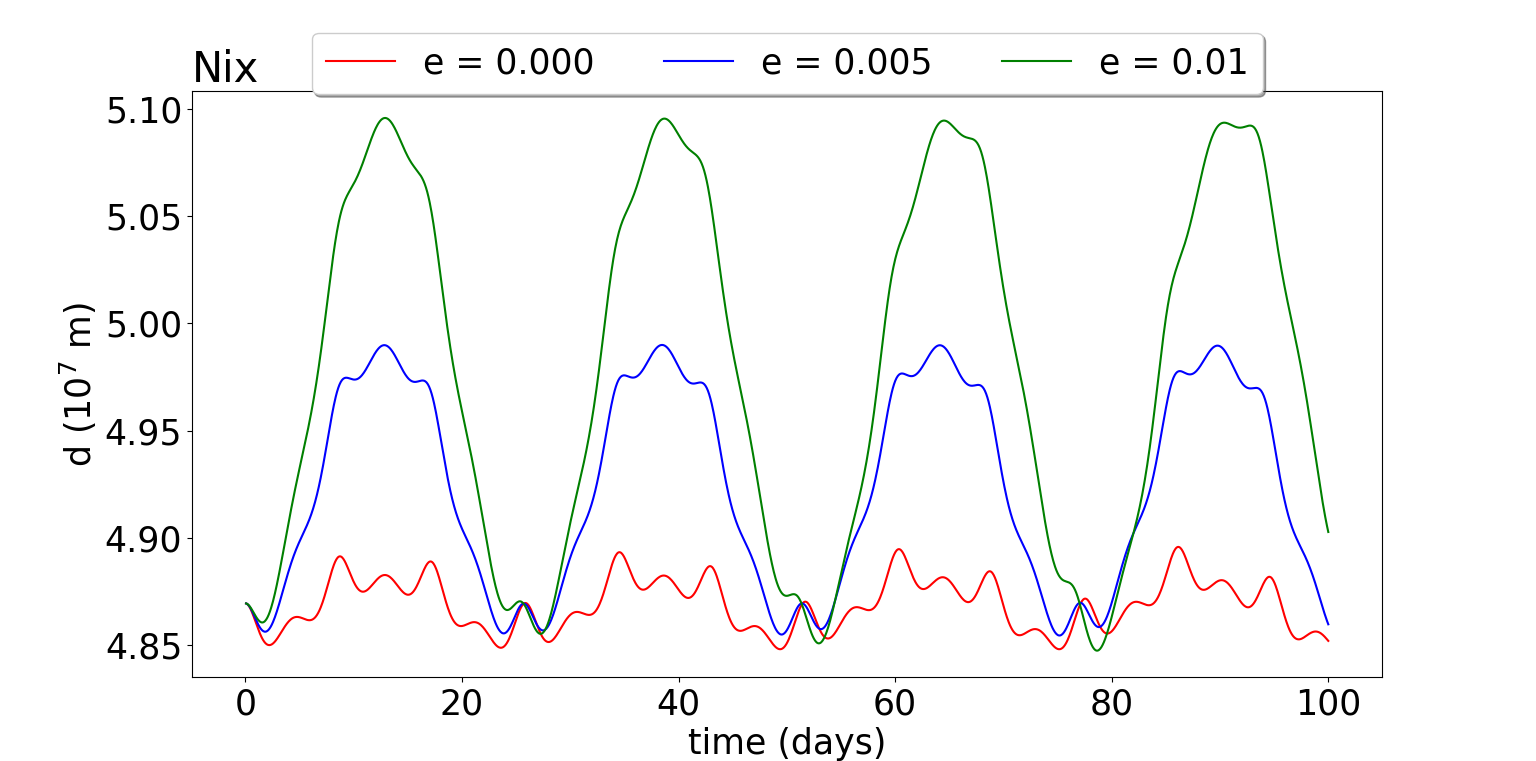} \\
     c\includegraphics[width=0.49\textwidth]{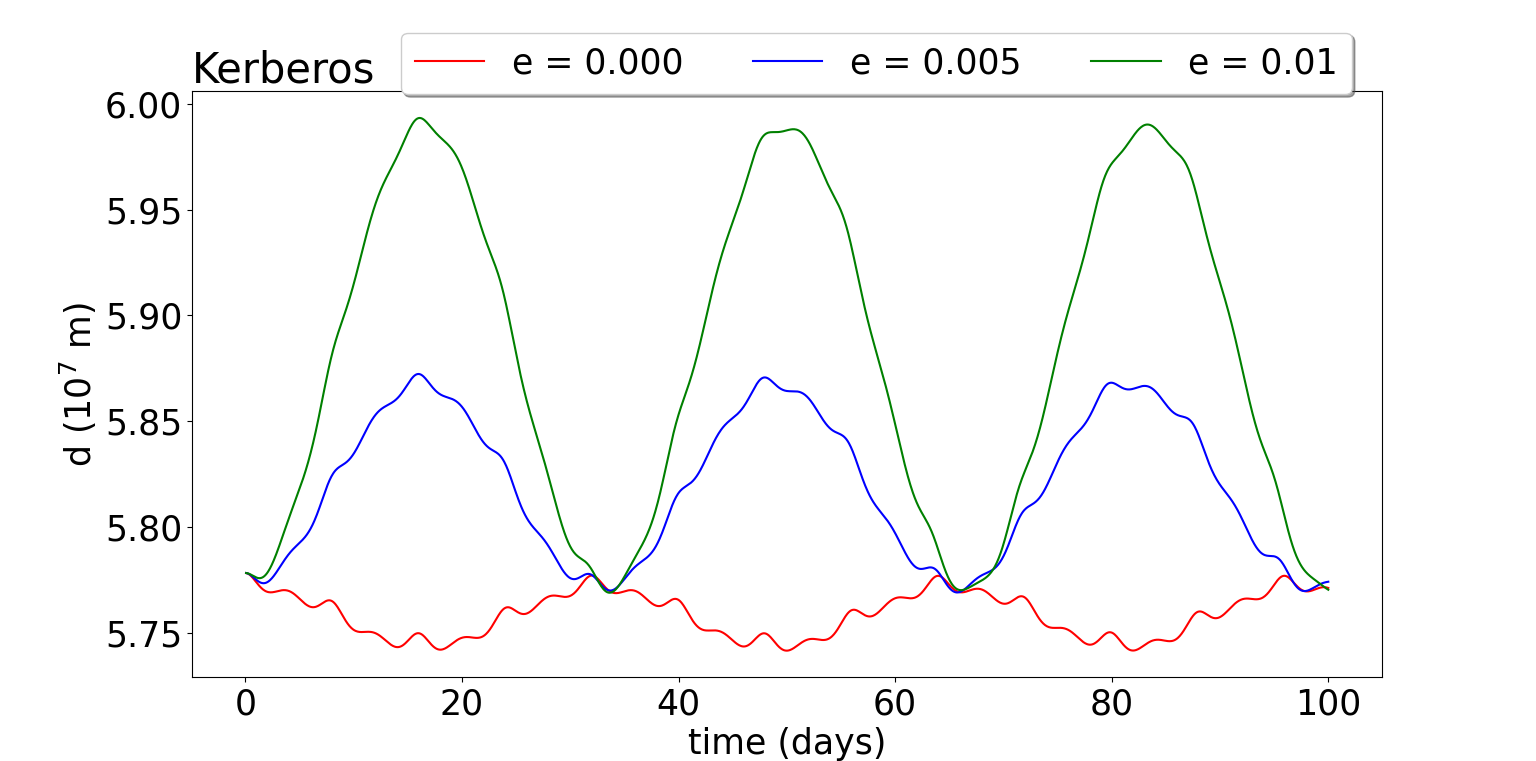} &
    d\includegraphics[width=0.49\textwidth]{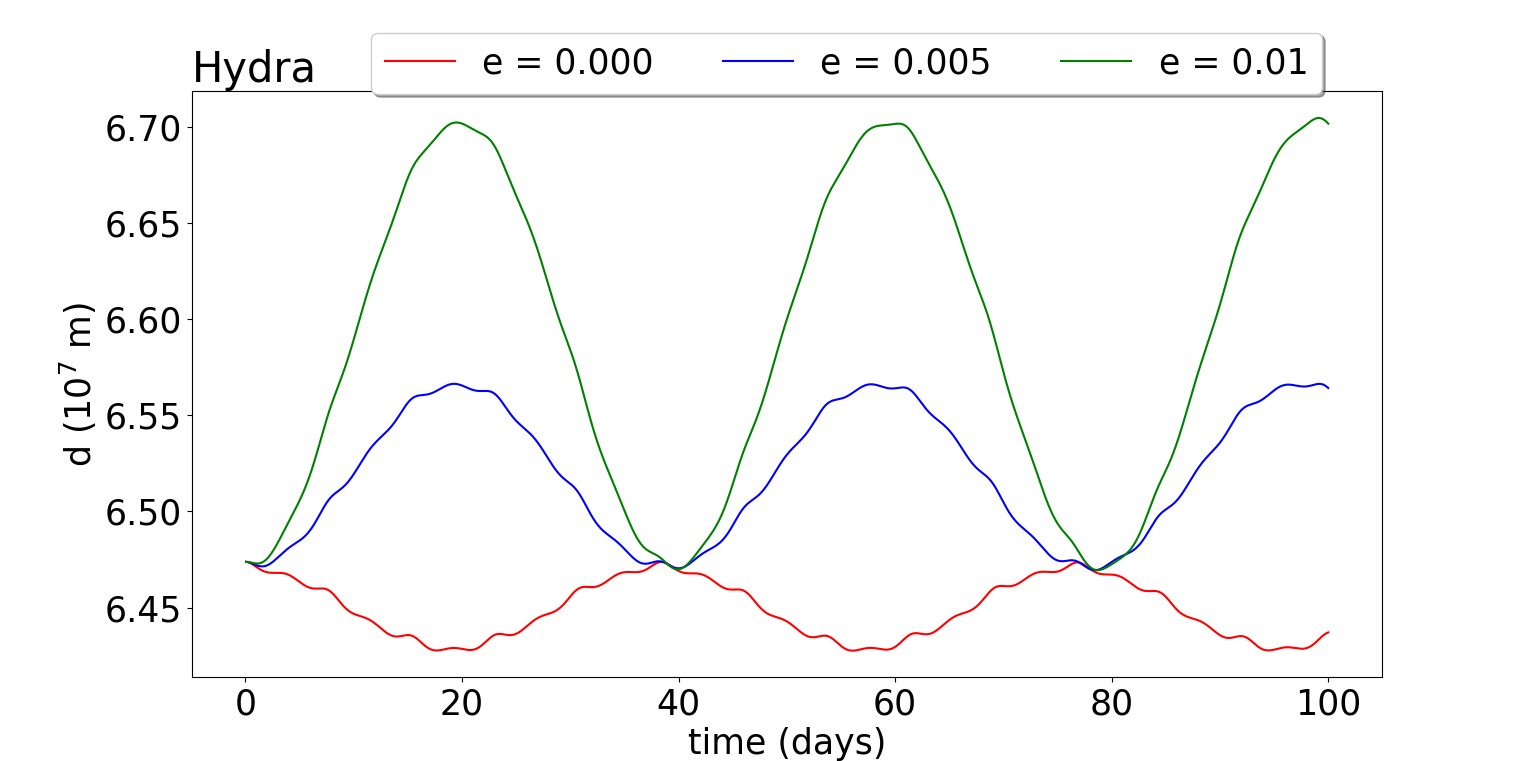} 
        \end{tabular} 
\caption{Distance from barycenter ($d$) through a timespan of 100 days by varying eccentricities. Red lines are for $e=0$, blue for $e=0.005$ and green for $e=0.01$.}
\label{fig:3}       
\end{figure*}

\section{Results and Discussion}
\label{sec:5}

Distance-from-barycenter plots of each object for each one of the aforementioned cases through a timespan of 5 years are shown in Figure \ref{fig:1}. Table \ref{tab:2} summarizes all different cases simulated. To get a better idea of the moon orbits on short timescales, we provide a zoom-in of the above plots for a total time of 100 days, which are a small multiple of orbital periods. These are shown in Figures 2a-2d. Each of these figures contains the obtained orbits of each small moon for cases A-F, as described in Section \ref{sec:4.2}.

Many fluctuations appear in all of the orbital distance plots, which arise from the superposition of larger or smaller oscillations of each object's motion.  Major oscillation periods are consistent with the moons' periods found in literature. Exact periods cannot be accurately defined for circumbinary orbits, as we explained in Section \ref{sec:3.1}. Smaller perturbations occur at timescales similar to the central bodies' period. These smaller oscillations don't appear to have a constant amplitude though. This behavior is expected, considering that the central potential caused by Pluto and Charon changes as time passes during their orbits. As a result, oscillations depend on the relative positions of the bodies every moment. 

In our simulations corresponding to cases B-E, we selected suitable velocities for each body, so that the resulting orbit has an eccentricity as close as possible to zero. As mentioned above, moons do not remain in their initial distances away from barycenter. Hence, this behavior is due to the time-dependent non-axisymmetric central potential and not some normal deviation from mean distance as it happens in usual elliptic orbits. Adopting slightly larger eccentricities results in even more high-amplitude motions. This is explicitly shown in Figures 3a-3d, which show how barycentric distances change by varying orbital eccentricity. Minimum-to-maximum orbital distance difference is almost doubled when we increase eccentricity from 0.005 to 0.01 for all moons. We also observe that when the eccentricity rises, the effect of the forced oscillations by the binary is weakened (less small-scale deviations are detected). This is of course expected by eq.~(4), where the second term, which corresponds to arbitrary motions as described by the eccentricity, dominates over the last one.

\begin{figure*}
\begin{tabular}{p{0.49\textwidth} p{0.5\textwidth}}
  a\includegraphics[width=0.49\textwidth]{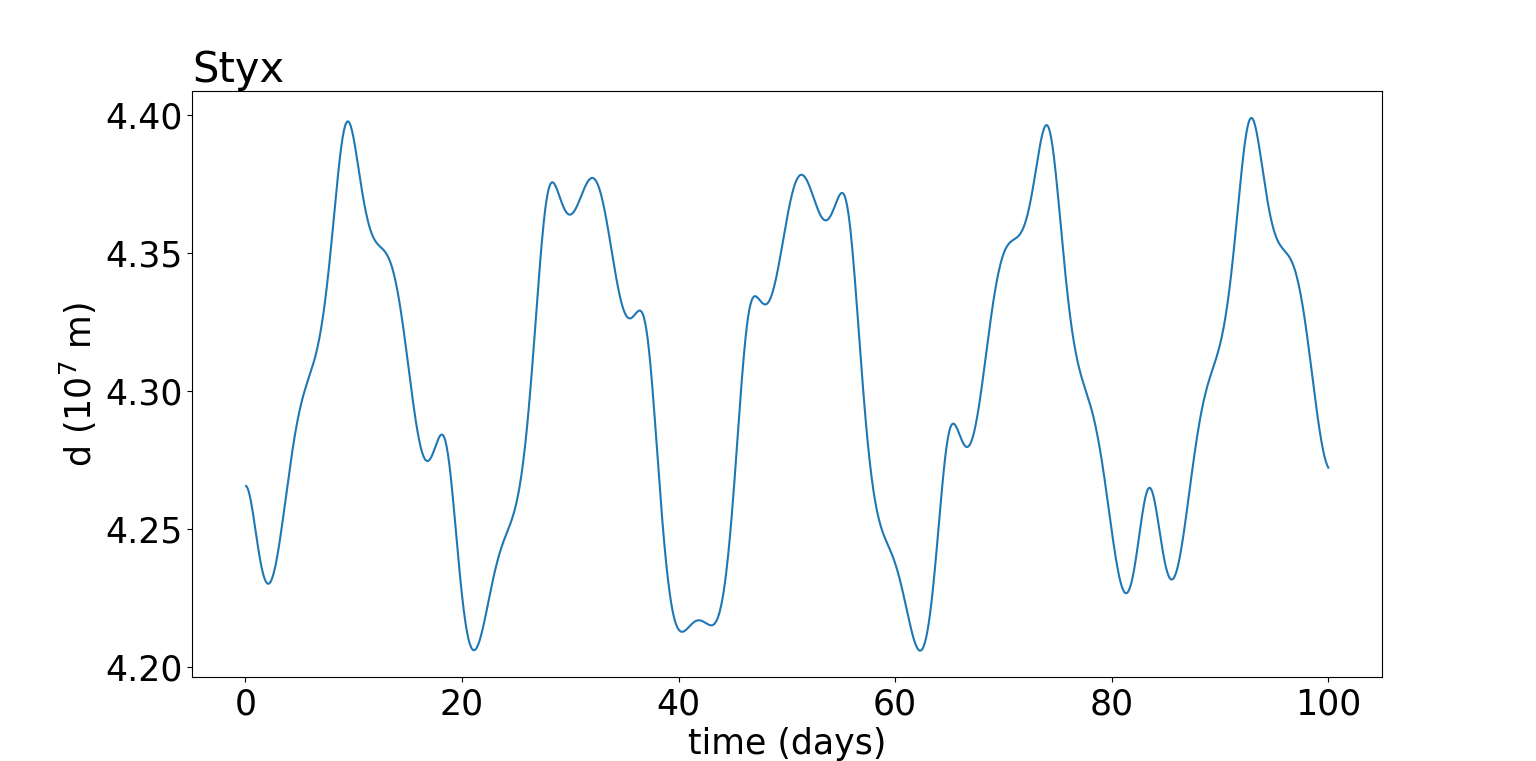} & 
    b\includegraphics[width=0.49\textwidth]{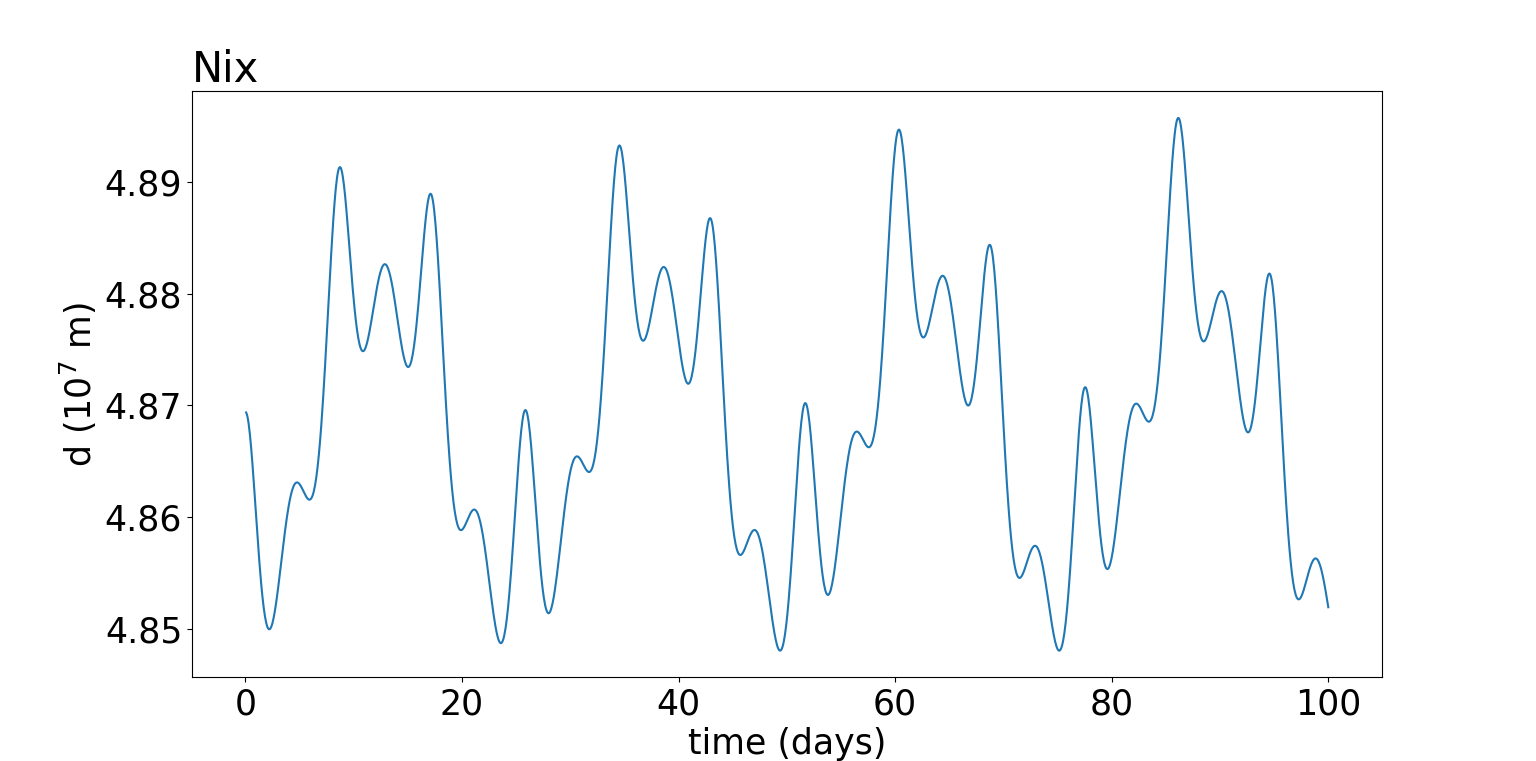} \\
     c\includegraphics[width=0.49\textwidth]{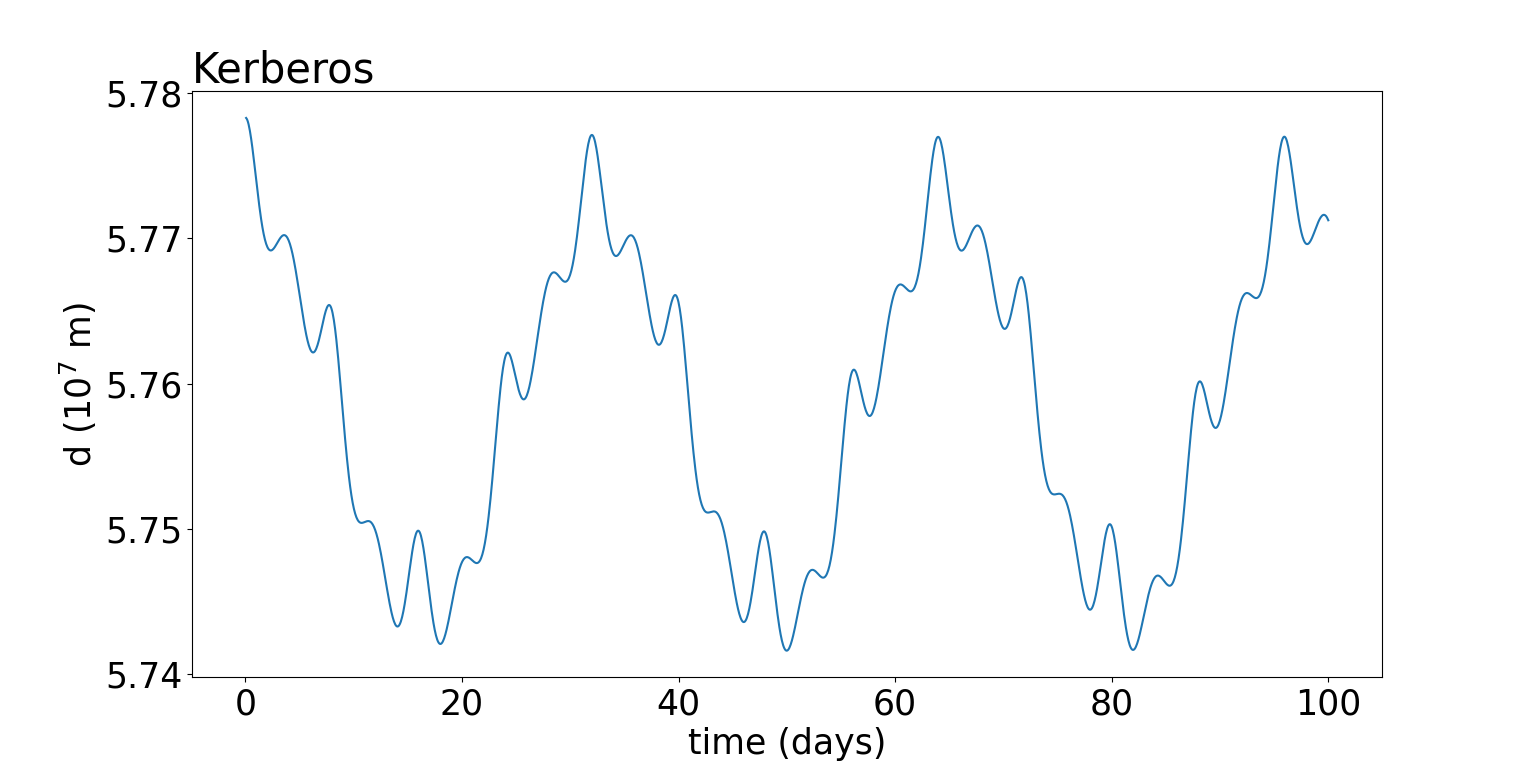} &
    d\includegraphics[width=0.49\textwidth]{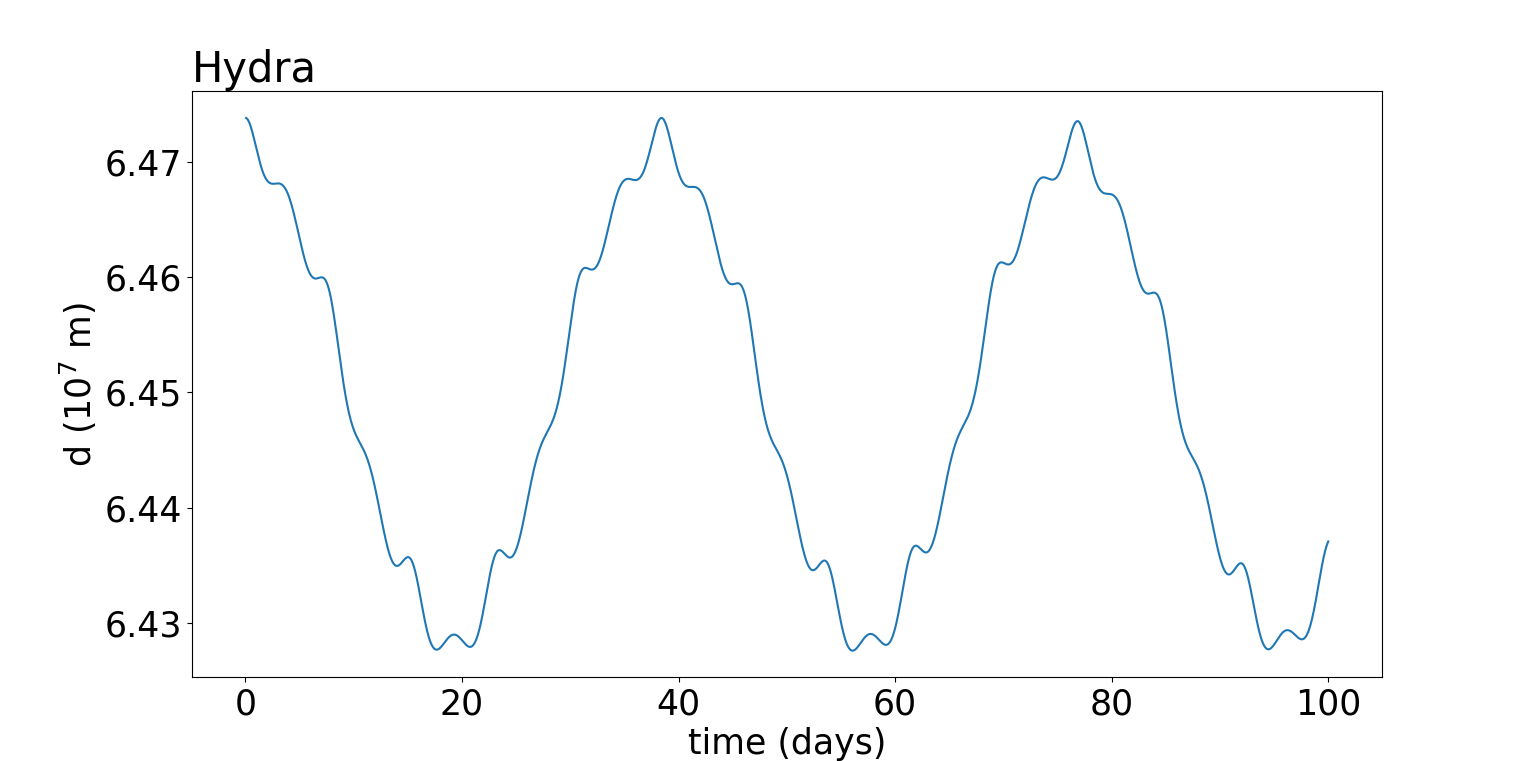} 
        \end{tabular} 
\caption{Variation of distance from barycenter ($d$) through a timespan of 100 days for most-circular orbits of Styx (panel a), Nix (panel b), Kerberos (panel c) and Hydra (panel d).} 
\label{fig:4}       
\end{figure*}

We confidently confirm that the relatively large variation in the distance of a moon from the barycenter is due to the central binary. To isolate the effect of Pluto and Charon to the moons' orbits, in some of our n-body integrations we combined their masses into a single object that replaced these two (case F). Having merged the binary into a single body in order to test Keplerian orbits, we observe that the deviations in distance are greatly shallower (dashed lines in Figures 1a-1f). These remaining perturbations are generated by intrinsic orbital eccentricities and the mutual gravitational attraction of the moons to each other. Yet, their effect should be minuscule when compared with the central binary's effect. Consequently, we determine that large deviations in the distance plot of a moon are at most caused by the time-varying non-axisymmetric central potential. No matter how hard one tries to enforce a minimal eccentricity on the orbits of these objects, the variation of their distance from the barycenter will bear a strong resemblance to an elliptical orbit. Specifically, orbital variations are only minimised when the moons are orbiting around a single object. 

Fluctuations become smaller for outer moons. Oscillatory terms drop as we move further away and gradually disappear, as the gravitational effect of Pluto and Charon diminishes at larger distances. The fact that the central binary effect is reduced as we move farther away from the barycenter can be seen in Figures 4a-4d. In these Figures, orbital distances for the four small moons over 100 days are presented. Starting vectors are such that moons follow "most-circular" orbits. It is shown in particular that Hydra appears to have smaller short-scale deviations than the other moons closer to the binary. For instance, Styx's distance maximum deviation is 4.4\%, whereas for the other small moons it is below 1.0\%. Furthermore, as discussed below, Styx actually seems to have a quite irregular orbit, not having a constant pattern in its distance from the barycenter. We expect that at even larger distances from Pluto and Charon, fluctuations almost disappear and only a nearly sine wave remains.

\begin{figure*}
\begin{tabular}{p{0.49\textwidth} p{0.5\textwidth}}
  a\includegraphics[width=0.49\textwidth]{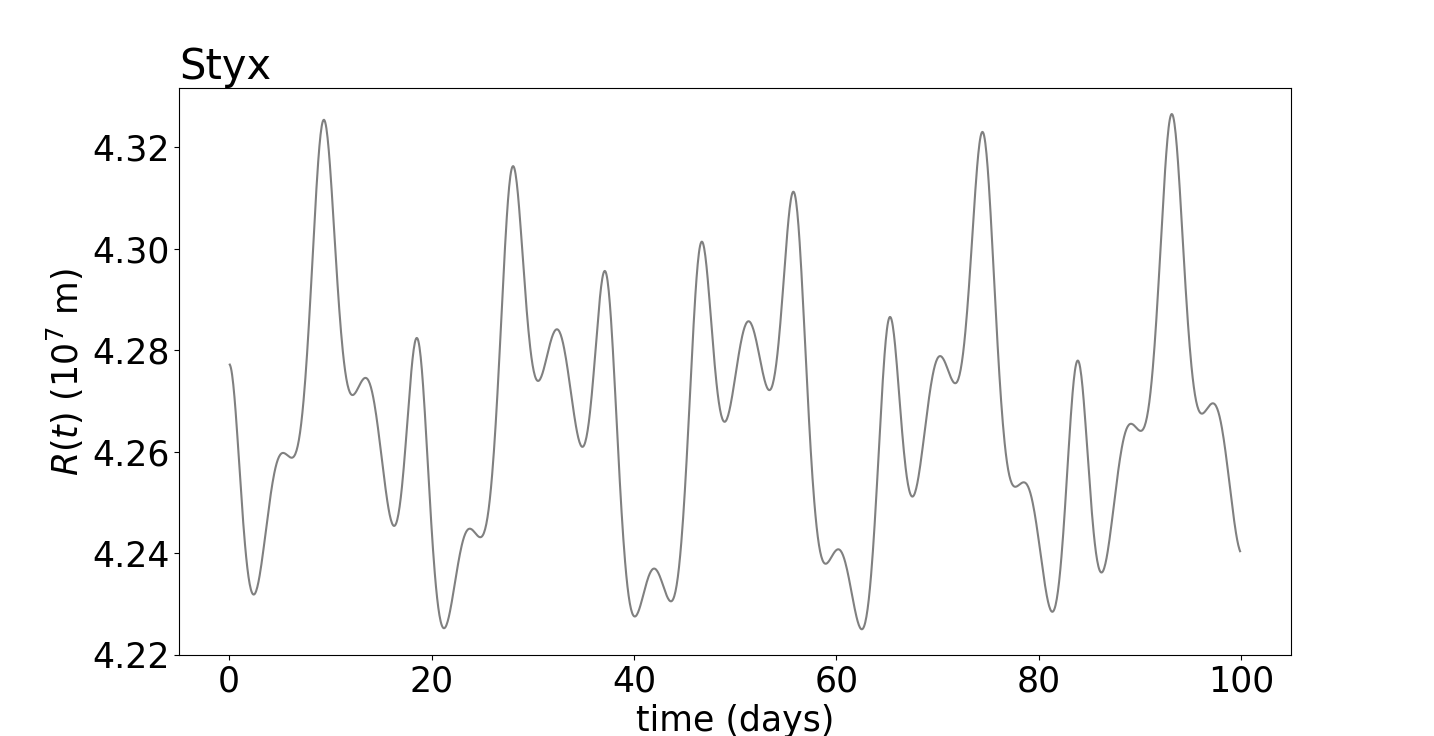} & 
    b\includegraphics[width=0.49\textwidth]{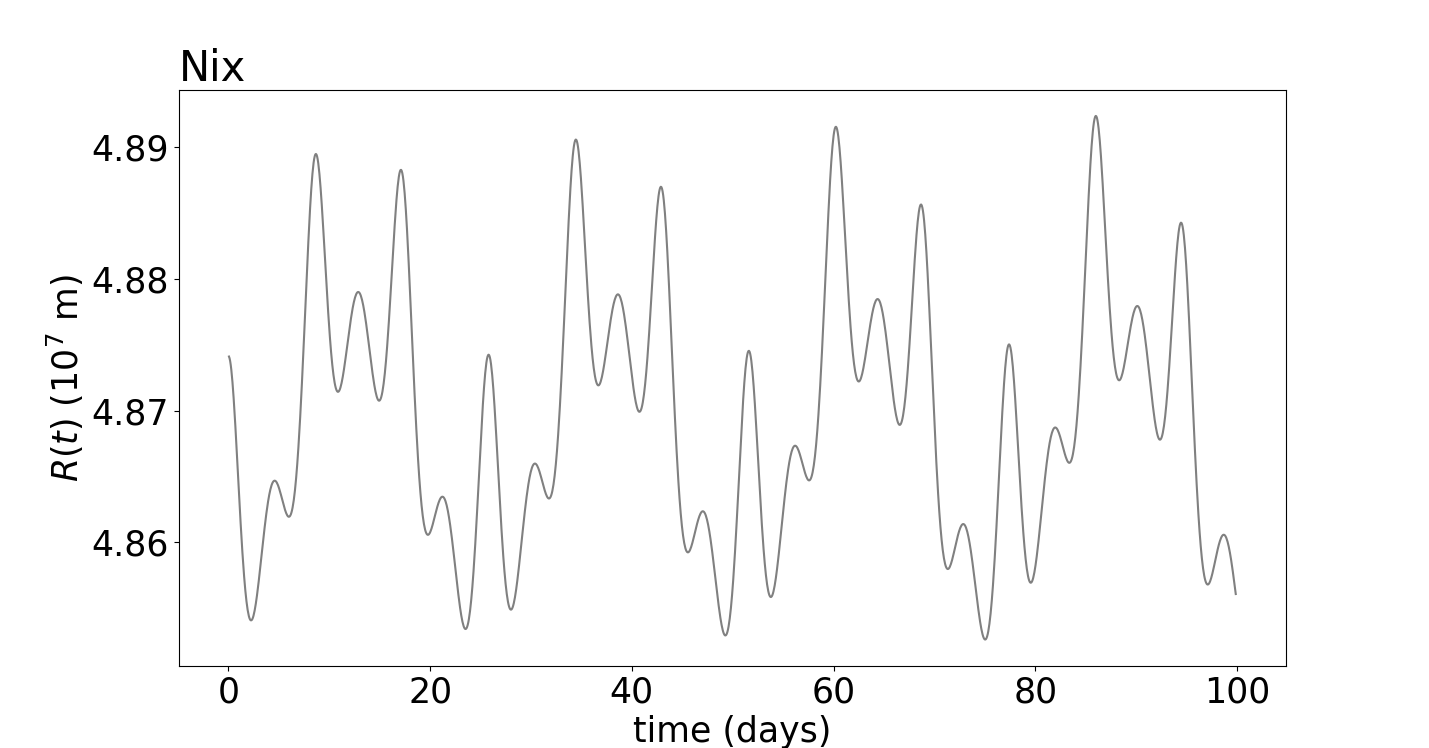} \\
     c\includegraphics[width=0.49\textwidth]{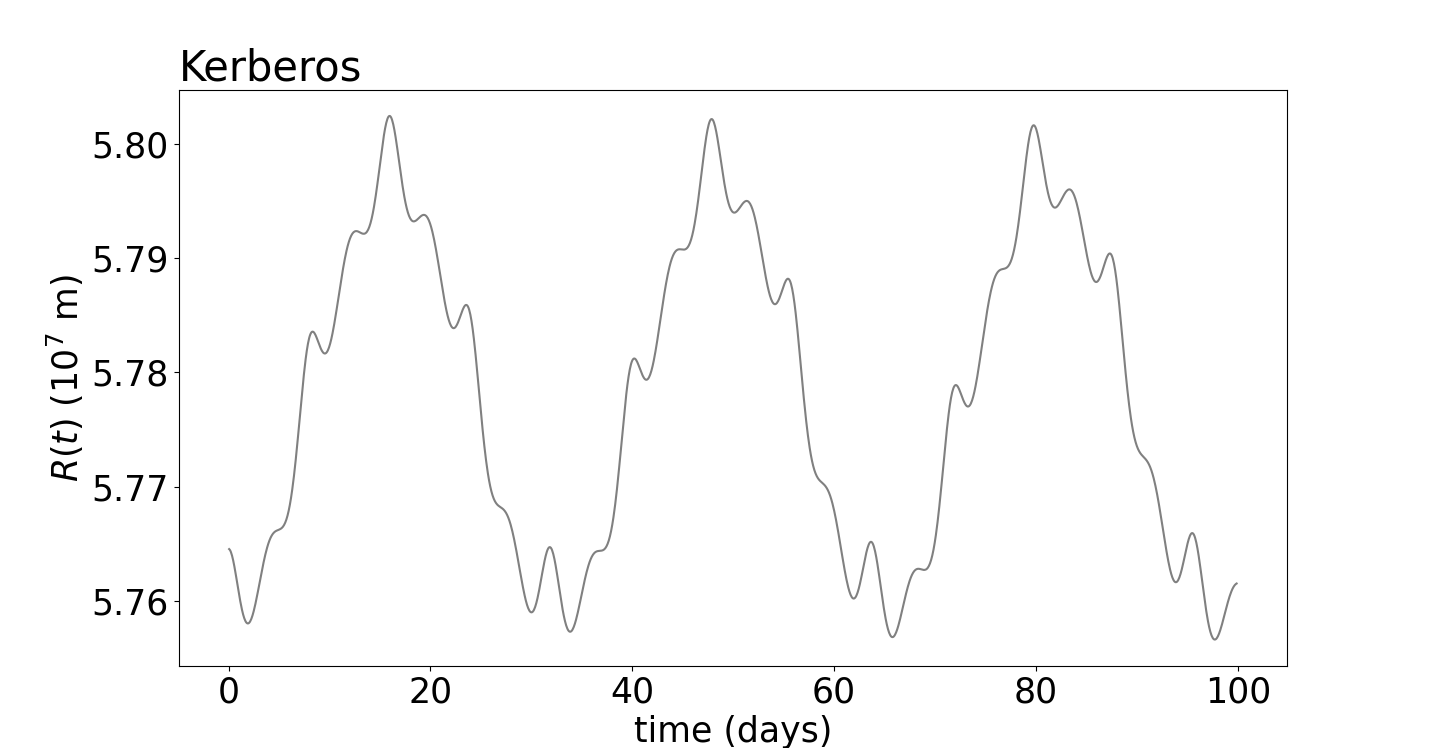} & 
    d\includegraphics[width=0.49\textwidth]{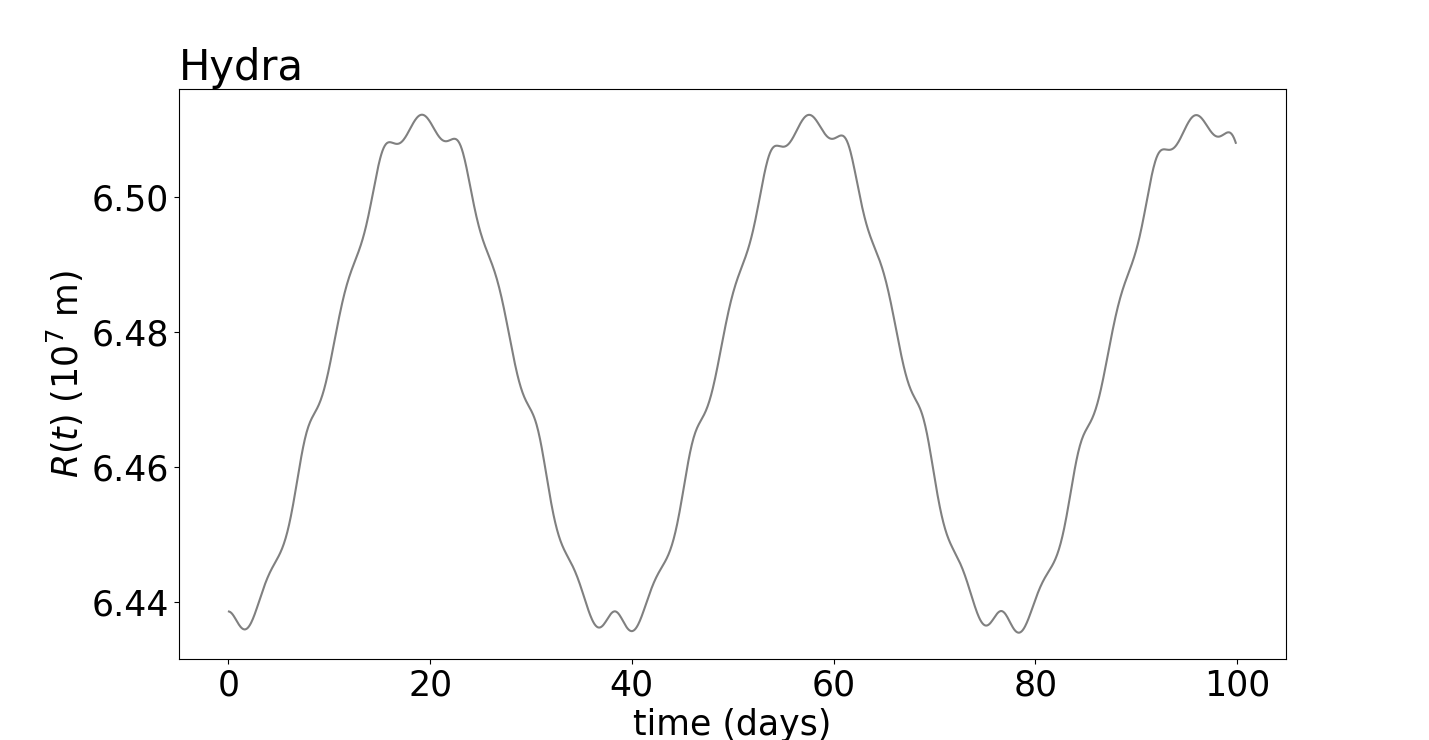} \\
        \end{tabular}
\caption{$R(t)$ evolution through a time period of 100 days for Styx (panel a), Nix (panel b), Kerberos (panel c) and Hydra (panel d) using the semi-analytic model. Free eccentricities for each moon were given the nominal eccentricities from Table \ref{tab:1}.}
\label{fig:5}       
\end{figure*}

\begin{figure*}
\begin{tabular}{p{0.49\textwidth} p{0.5\textwidth}}
  a\includegraphics[width=0.49\textwidth]{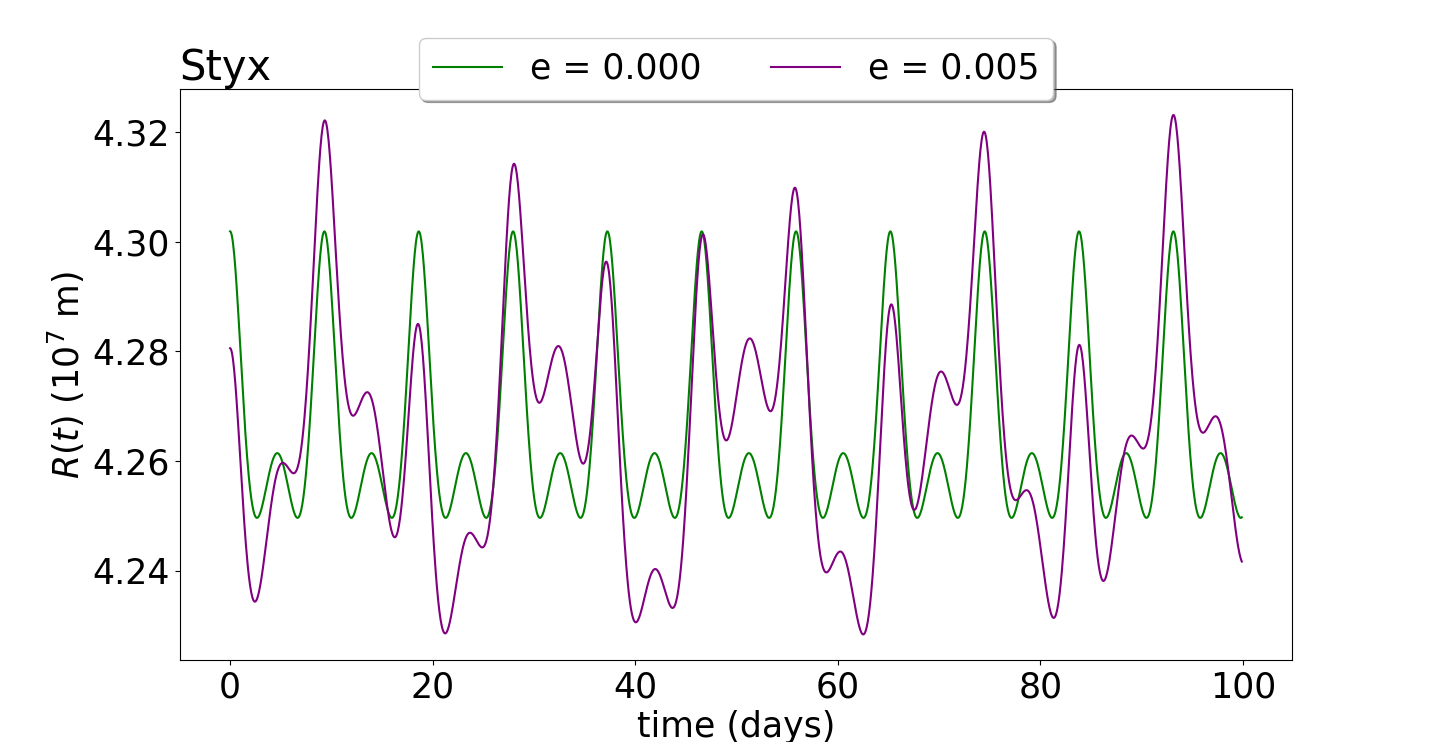} & 
    b\includegraphics[width=0.49\textwidth]{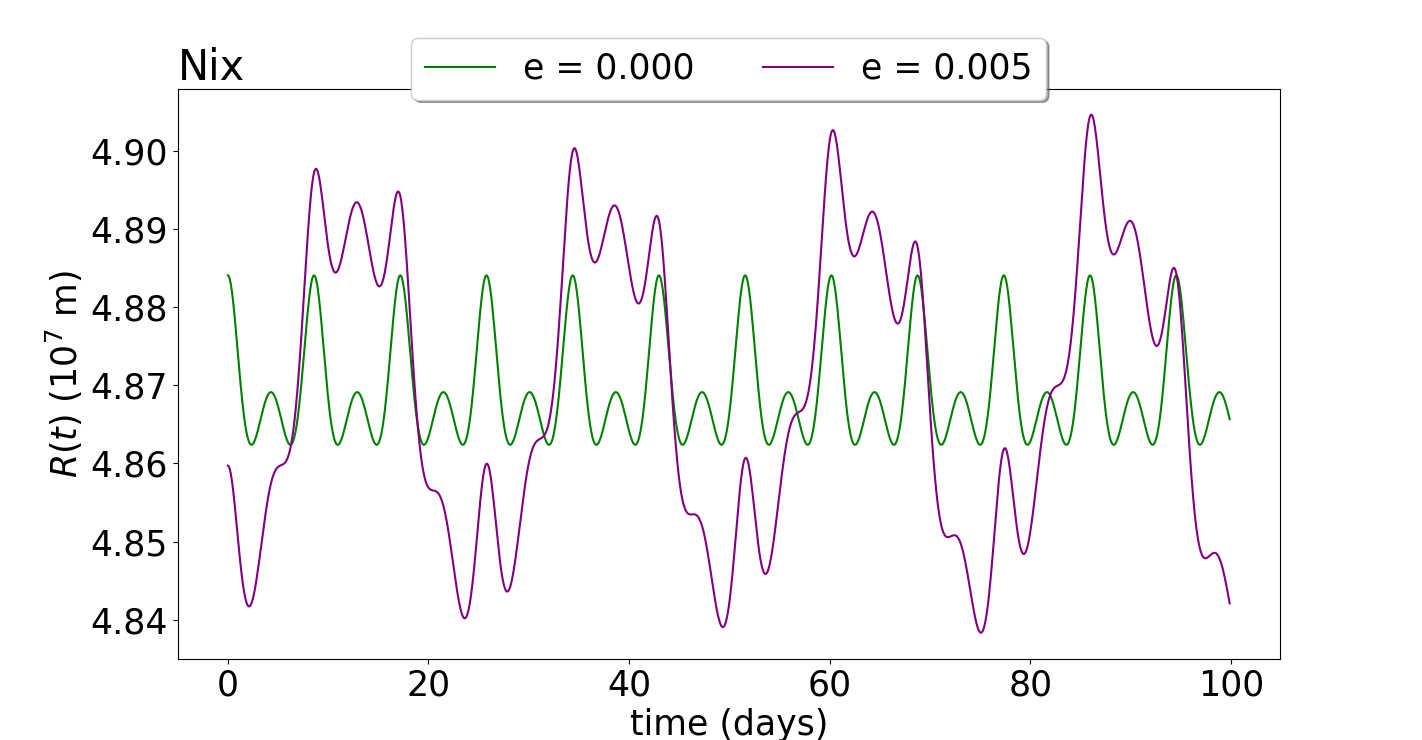} \\
     c\includegraphics[width=0.49\textwidth]{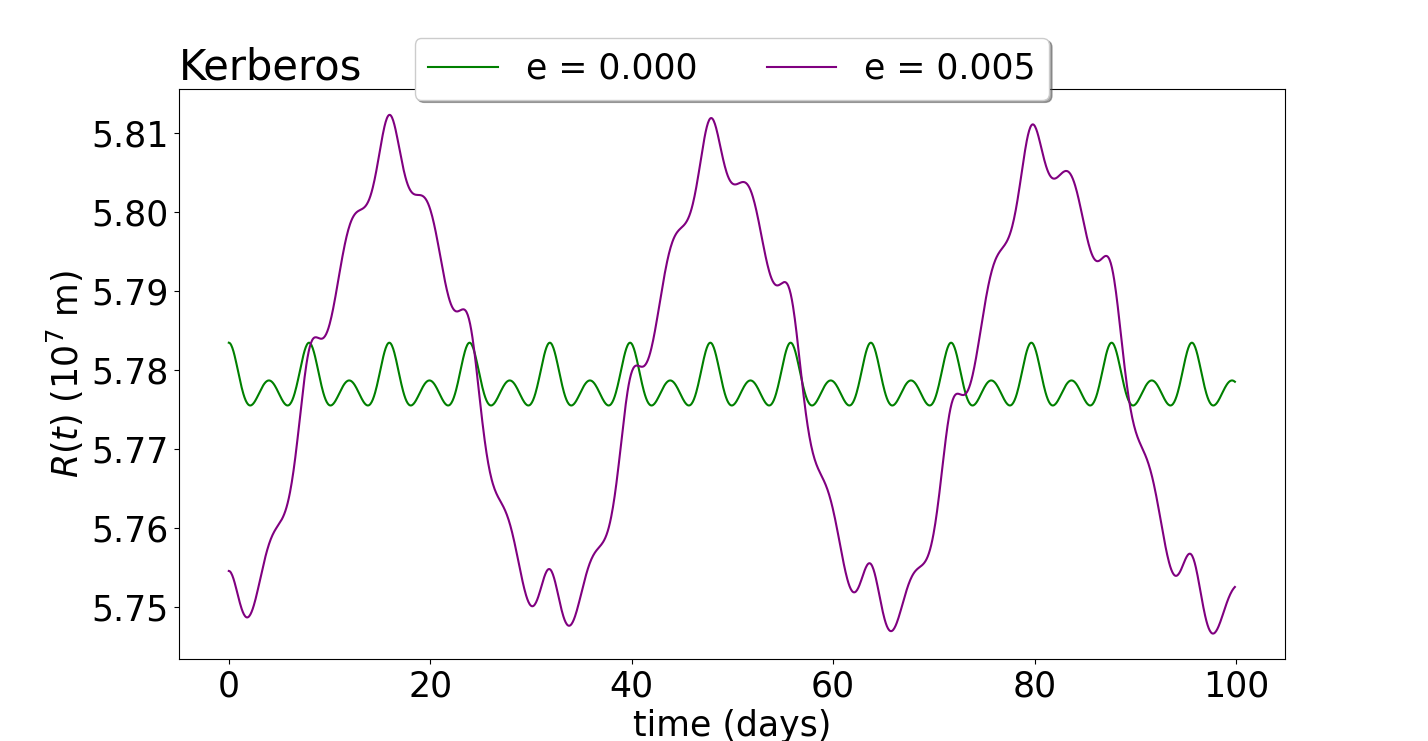} & 
    d\includegraphics[width=0.49\textwidth]{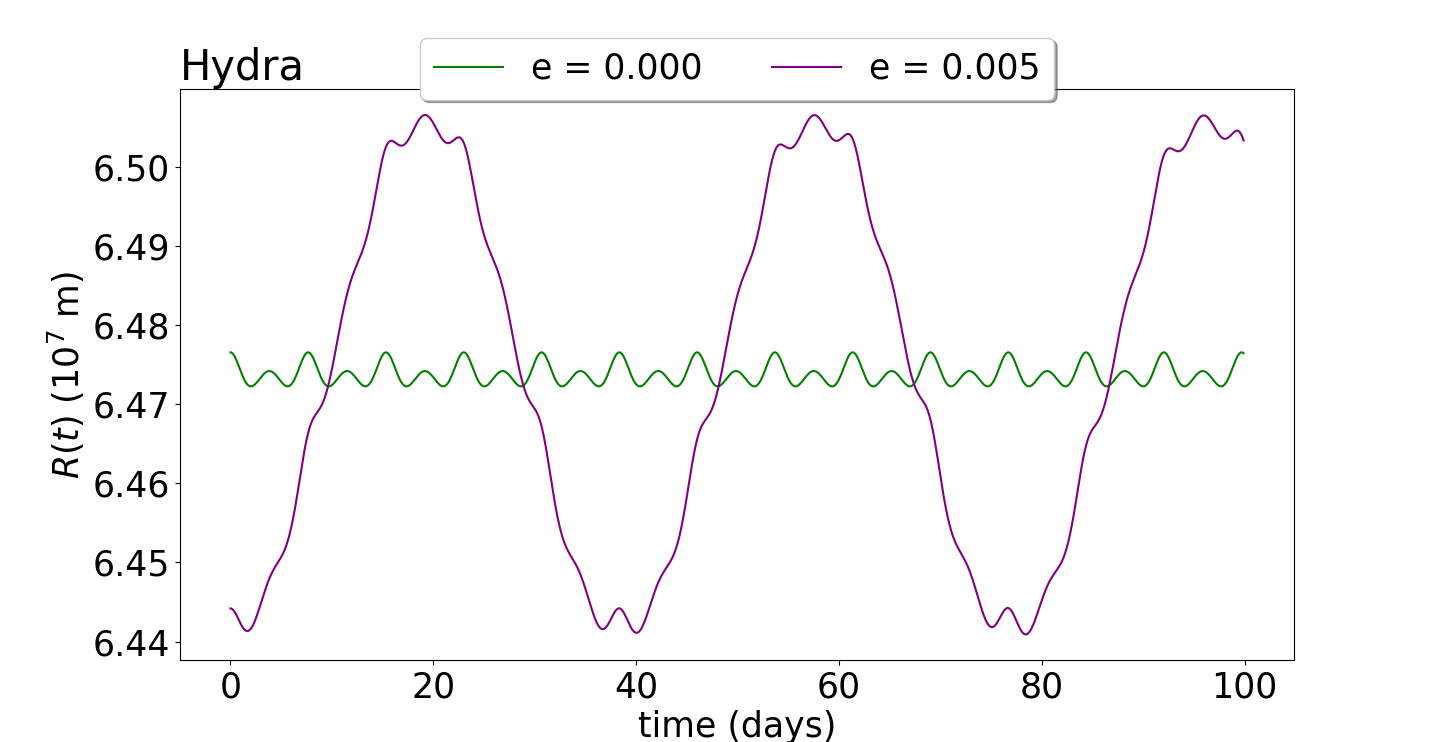} \\
        \end{tabular}
\caption{$R(t)$ evolution through a time period of 100 days for Styx (panel a), Nix (panel b), Kerberos (panel c) and Hydra (panel d) using the semi-analytic model by varying free eccentricities. Green lines are for $e=0.000$ and purple for $e=0.005$.}
\label{fig:6}       
\end{figure*}

We infer that the simulated orbits are quite similar to what Lee-Peale theory predicts. Orbits indeed seem to be a superposition of the circular motion of a guiding center, an epicyclic motion described by some "eccentricity" (free eccentricity) and random motion from Pluto and Charon. We provide for comparison distance plots for all small moons using the semi-analytic approach illustrated in Section \ref{sec:3.1} (Figures \ref{fig:5} and \ref{fig:6}). Distance plots in Figure \ref{fig:5} were obtained by setting each moon's nominal eccentricity (Table \ref{tab:1}) as the value of the free eccentricity parameter of eq.~(4). Accordingly, free eccentricities were given the values $e=0.000$ (green lines) and $e=0.005$ (purple lines) in Figure \ref{fig:6}. Again, as observed in the n-body-acquired plots, the growth of the free eccentricity results in higher-amplitude orbits, and simultaneously in the dominance of this factor over the central binary's effect. As expected, purple and green lines are not as distinct for Styx as they are for Kerberos or Hydra. This is explained by the proximity of Styx to the binary, and hence its larger dependence on the positions of Pluto and Charon rather than its elemental epicyclic motion. Ultimately, a non-zero value for free eccentricity gives more irregular orbital patterns, as an additional cosine term contributes in the solution. 

\begin{figure*}
\begin{tabular}{p{0.49\textwidth} p{0.5\textwidth}}
  a\includegraphics[width=0.49\textwidth]{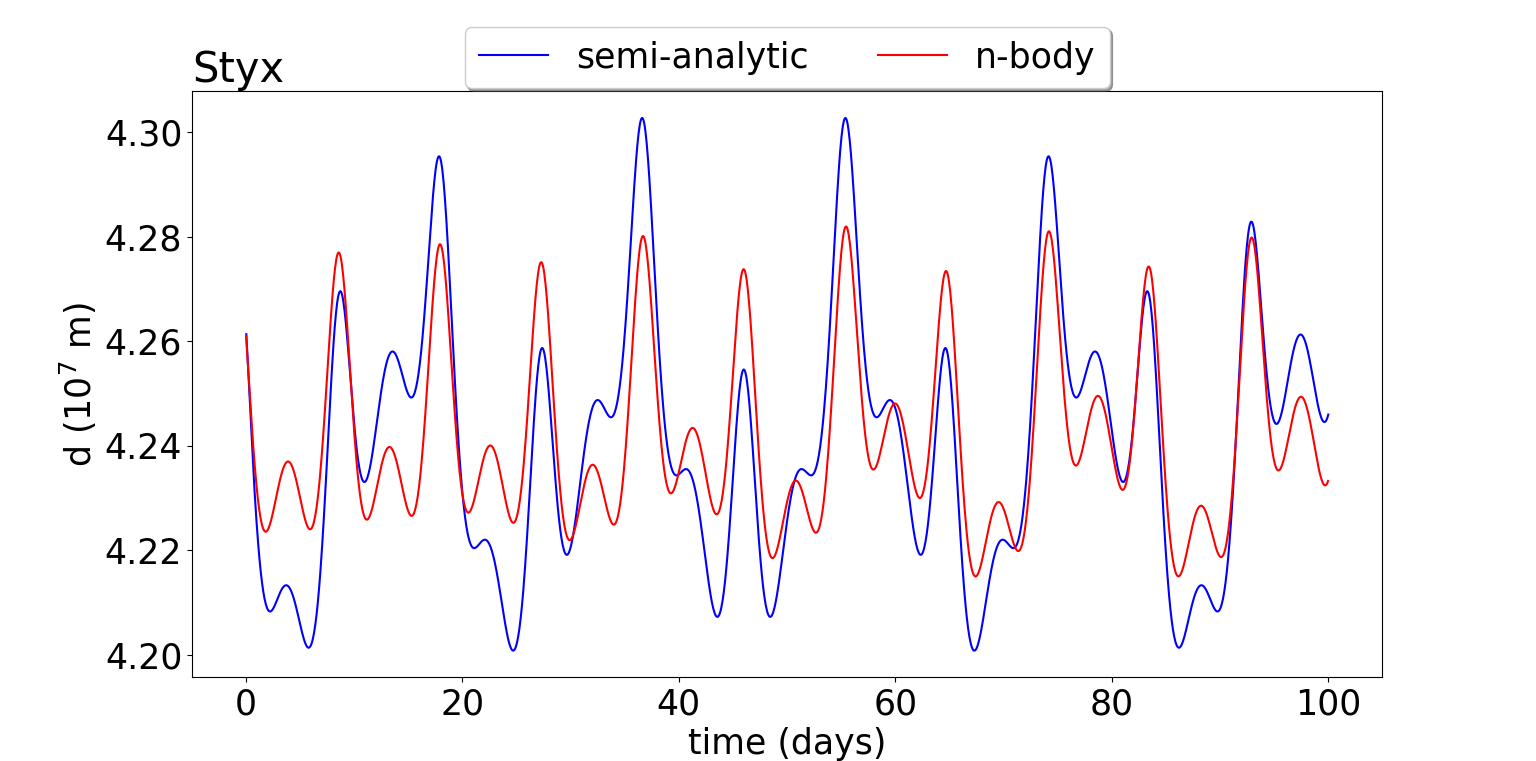} & 
    b\includegraphics[width=0.49\textwidth]{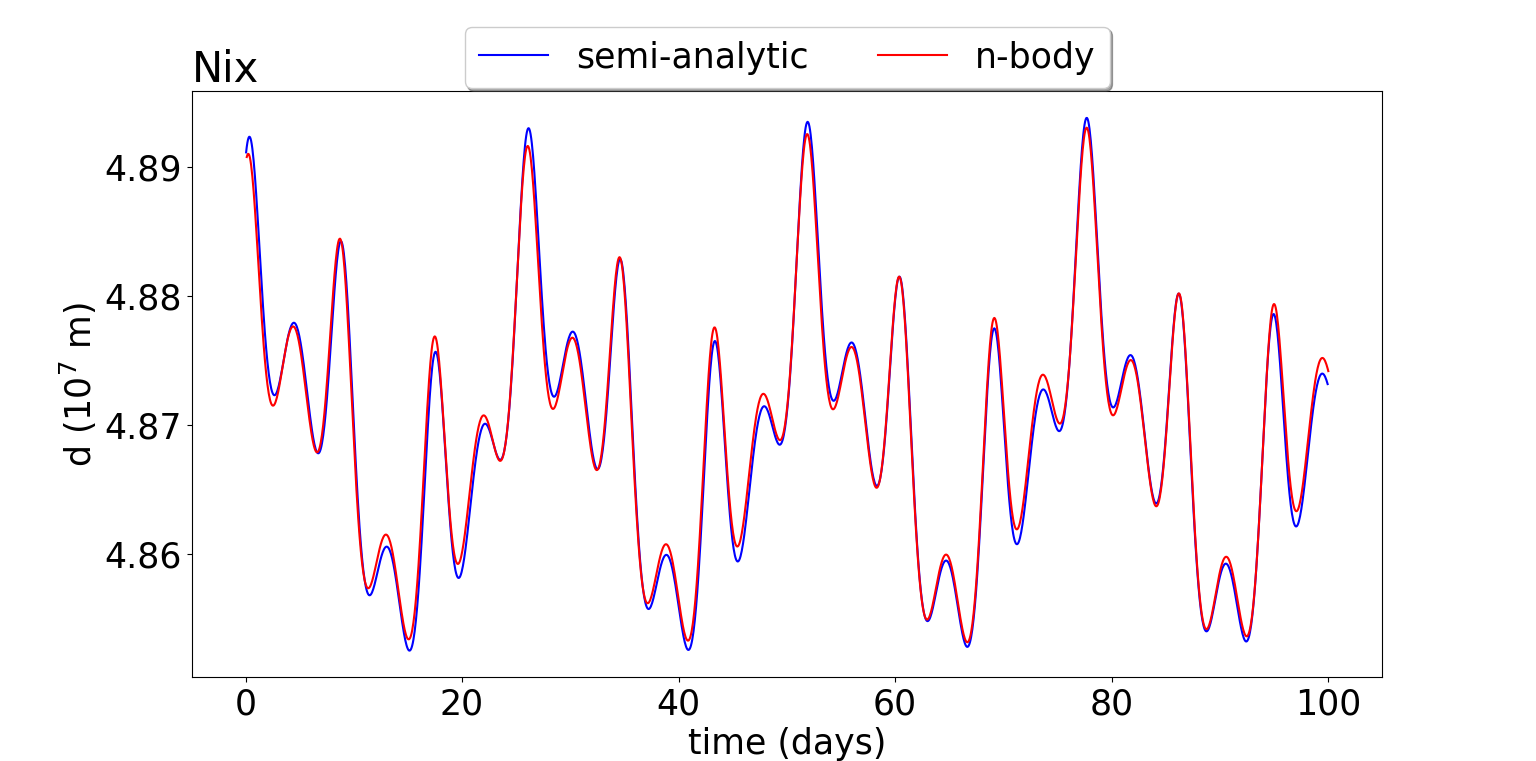} \\
     c\includegraphics[width=0.49\textwidth]{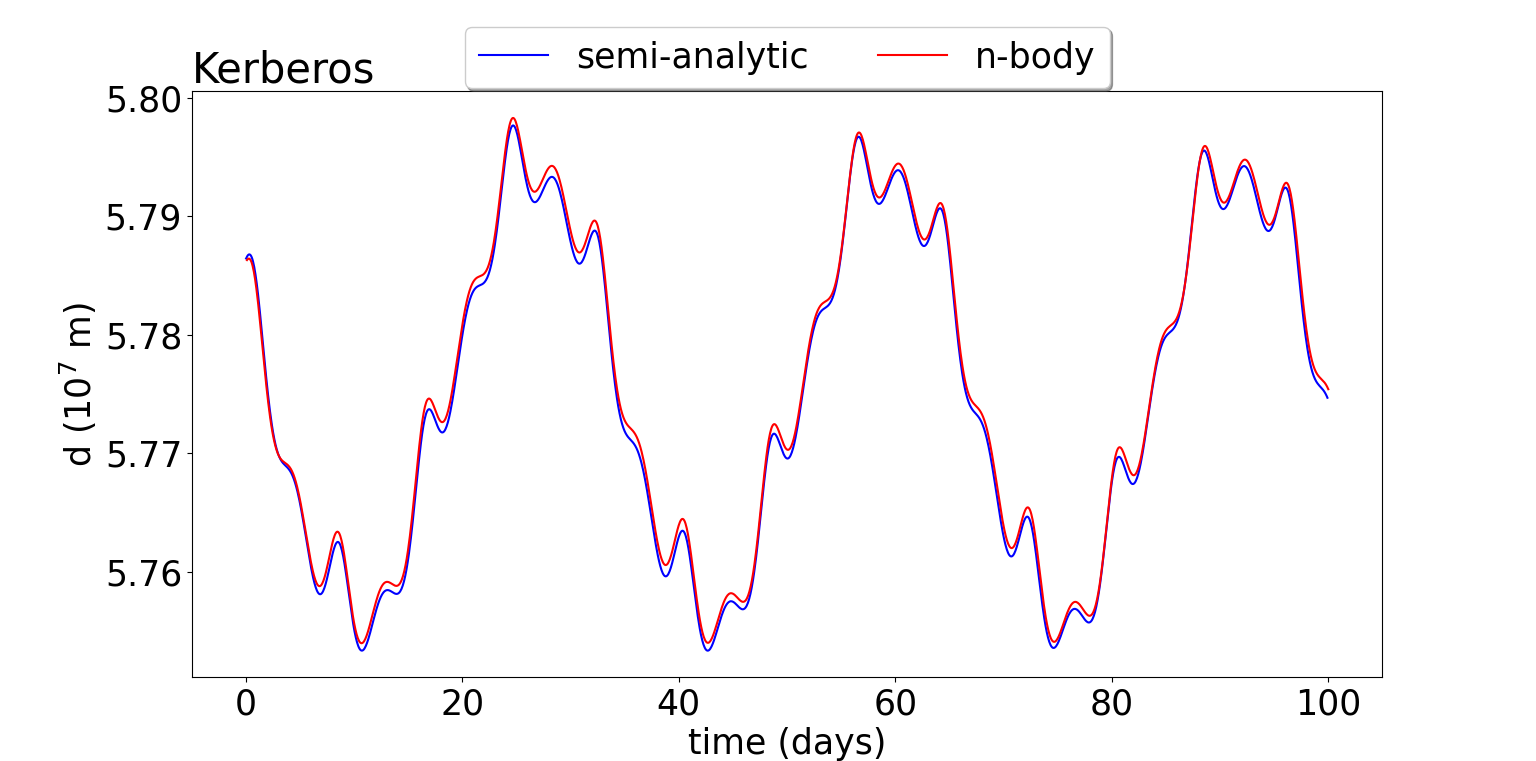} &
    d\includegraphics[width=0.49\textwidth]{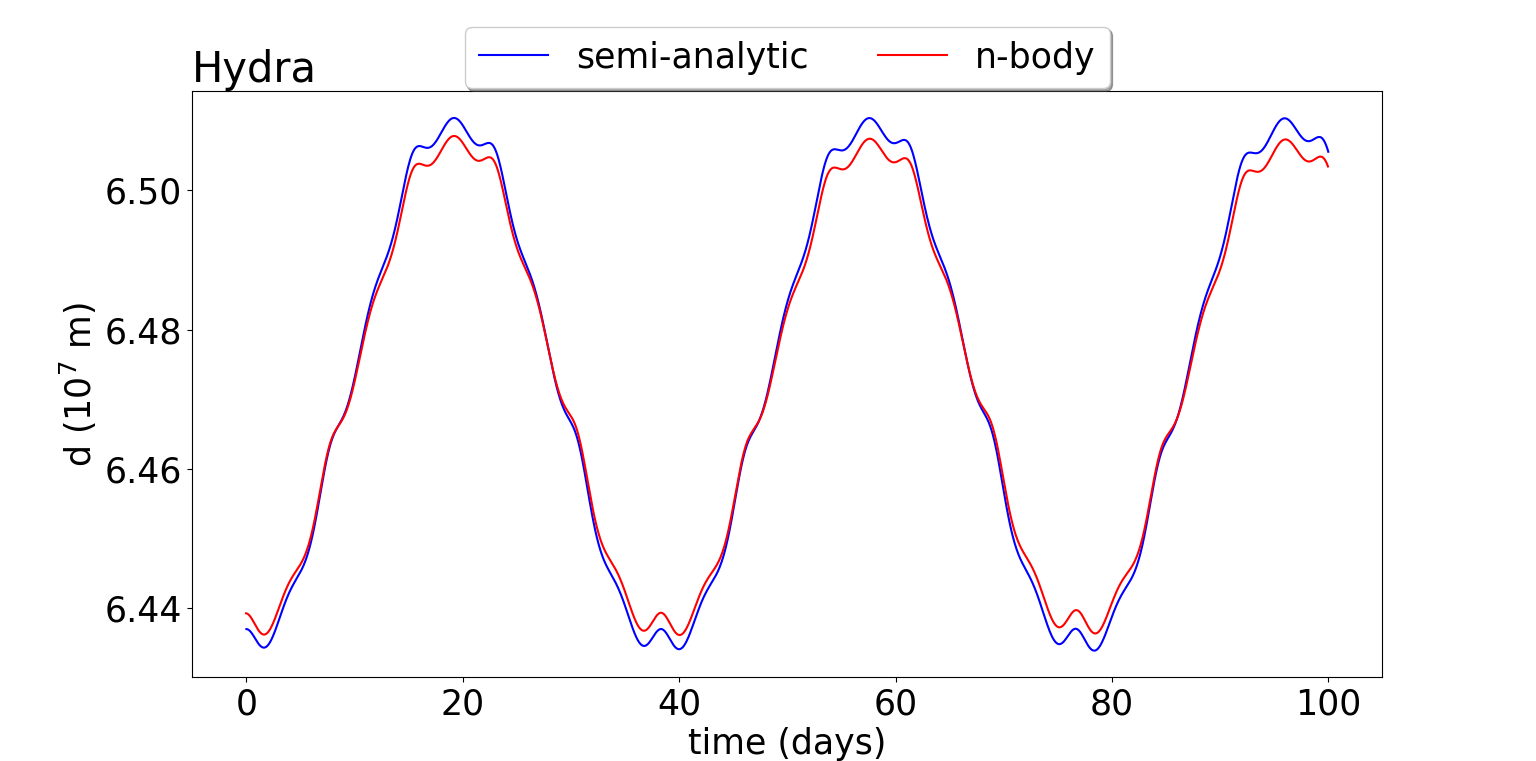} \\
        \end{tabular}
\caption{Comparison of the variation of distance from barycenter ($d$) through a timespan of 100 days  for Styx (panel a), Nix (panel b), Kerberos (panel c) and Hydra (panel d) using the n-body simulation and the semi-analytic model.}
\label{fig:7}       
\end{figure*}

In order to calculate the relative difference between the values obtained by the n-body integrator, $R_{m}$, and the the semi-analytic method, as specified by eq.~(4), $R_{n}$, we define the following norm: 
\begin{eqnarray}
\|\delta R \|=\frac{1}{(t_f-t_i)}   \int_{t_i}^{t_f} \frac{| R_{n} - R_{m} |}{\langle R \rangle}  \,dt \,,
\end{eqnarray}
where $\langle R \rangle$ is the mean value deduced by the two approaches and $t_f - t_i $ is the simulated timespan. Comparison plots for the two methods are given in Figure \ref{fig:7}. Red lines represent the variation of orbital distance through time for the case A (see Section \ref{sec:4.2}) of n-body simulations, whereas the blue ones are the depiction of eq.~(4), by assigning nominal eccentricities as free eccentricities. The results of $\|\delta R \|$ show that the deviation of the two methods is 0.2925\% as for Styx, 0.0192\% for Nix, 0.0117\% for Kerberos and 0.0267\% for Hydra. It is verified then that the linearized theory by \cite{Lee:2006} is adequately accurate to describe circumbinary orbits. A substantial part of the discrepancy between the two methods is induced by the fact that only the n-body simulations contain the entire system, and therefore the additional interactions of every moon to each other. More precisely, note that Styx is the lightest body, hence it is more intensely affected by gravitational perturbations caused the other moons. As a result, the deviations of the n-body code from semi-analytic model are more obvious for Styx. 

Moreover, a careful calculation of Kepler's 3rd law would suggest that values for orbital periods and semi-major axes taken by the specified papers, \cite{Brozovic:2015} and \cite{Showalter:2015} are not consistent. It is of course unsurprising, as a rotating time-changing potential alters significantly orbits and diverges them from what Kepler's laws indicate. Consequently, we argue that determined uncertainties in these two papers do not emerge because of incomplete or erroneous data. While this is certainly an important factor and more observations are needed, these inconsistencies are mostly caused from the central binary. In other words, they are intrinsic differences in the system's positions and do not actually reflect major systematic or other errors in the measurements.

We also note that osculating semi-major axes of Styx are quite different between the above two studies. In \cite{Brozovic:2015}, $a=42,413$ km was derived for Styx, whereas in \cite{Showalter:2015} the corresponding value is $a=42,656$ km. The discrepancy of these values is $\sim 3\sigma$ given the uncertainly of $\pm 78$ km reported in \cite{Showalter:2015}. We further note that in the extended data set presented in \cite{Showalter:2015}, (Extended Data Table 1) the semi-major axis of Styx's considering data from 2006 to 2012, is $42,484\pm 82$ km, which has a $2\sigma$ deviation from the primarily value mentioned in the same work. These deviations can be indicative of the oscillation of the barycentric distance of Styx reported here.

\begin{figure*}
\begin{tabular}{p{0.49\textwidth} p{0.5\textwidth}}
  a\includegraphics[width=0.49\textwidth]{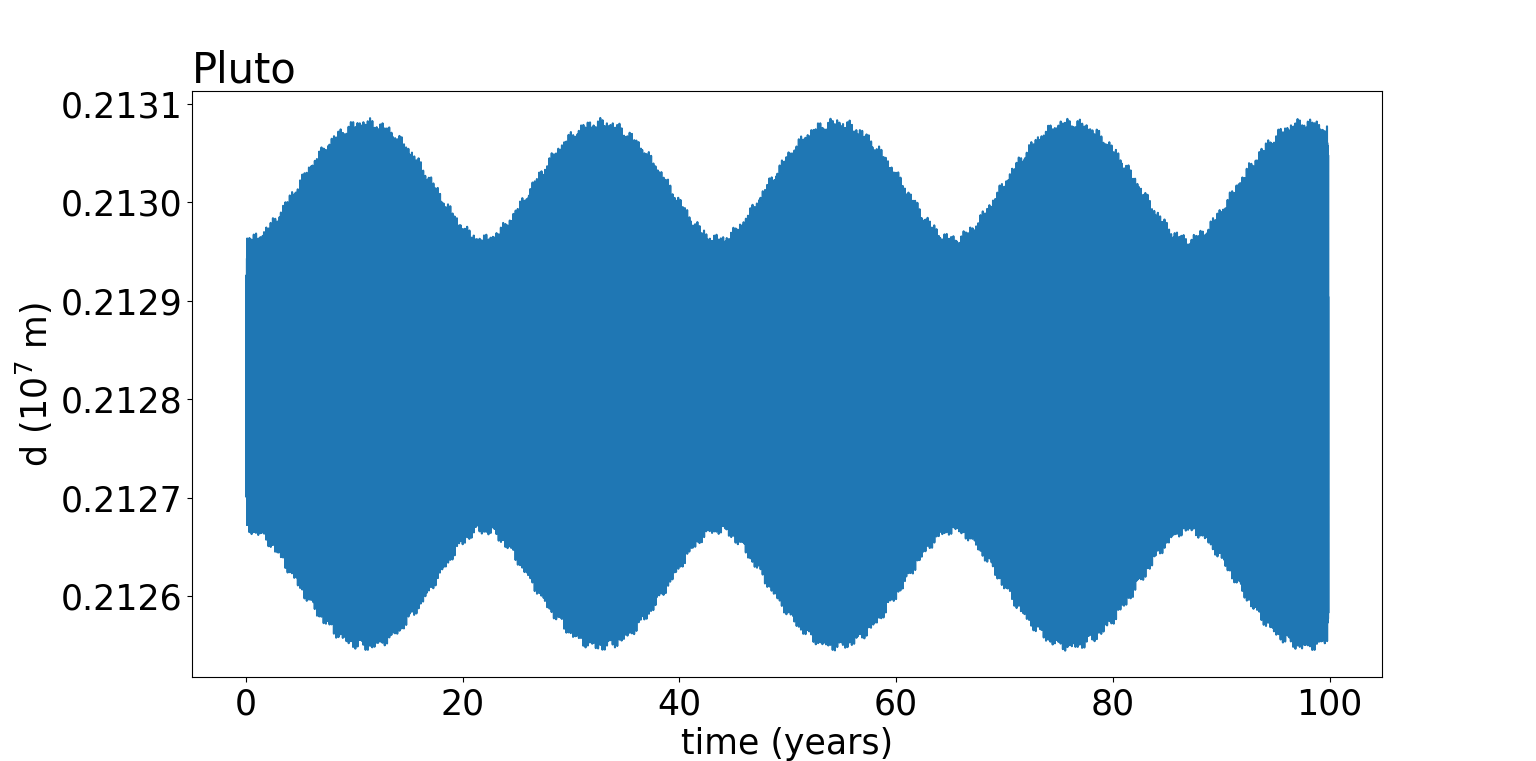} & 
    b\includegraphics[width=0.49\textwidth]{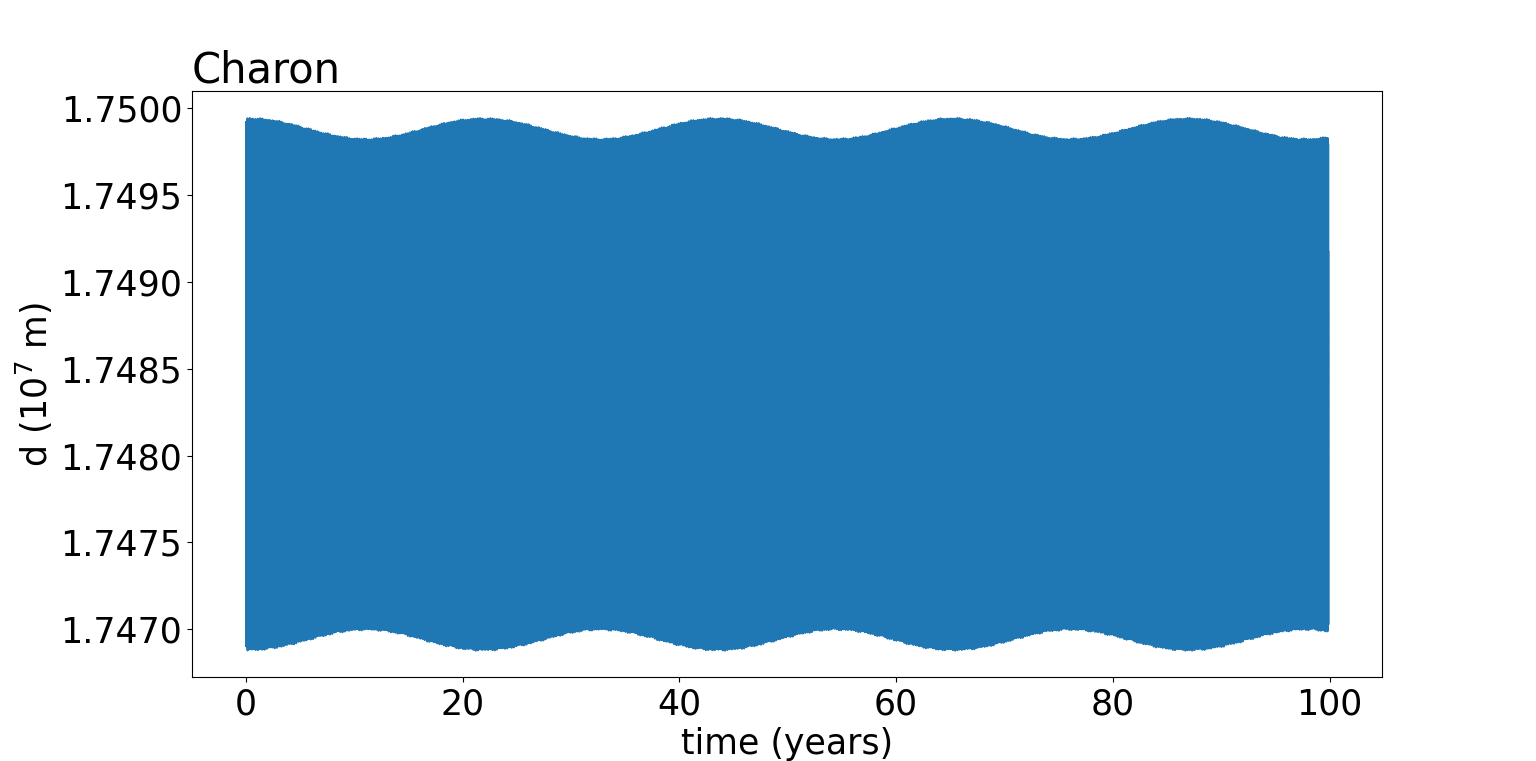} \\
     c\includegraphics[width=0.49\textwidth]{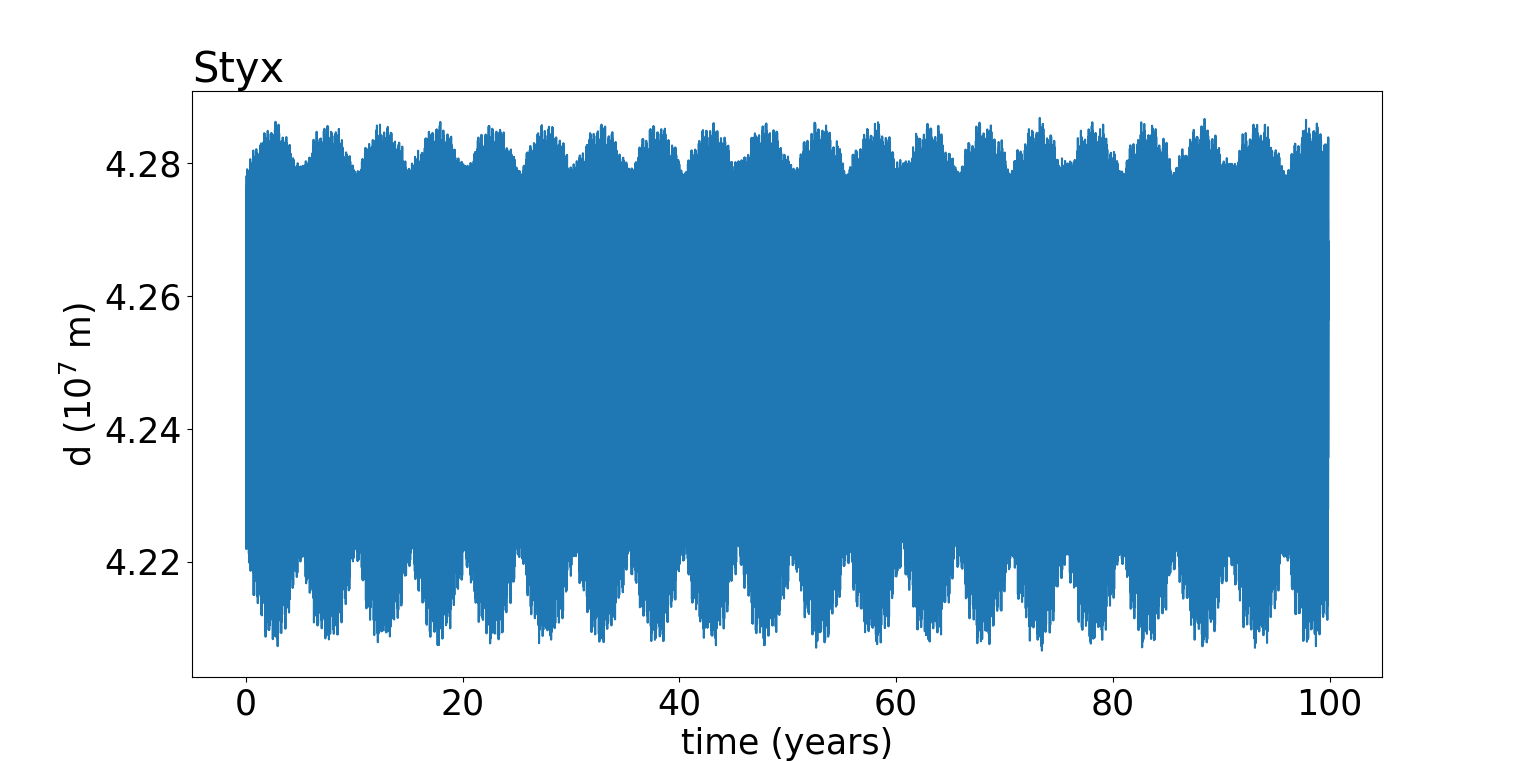} &
    d\includegraphics[width=0.49\textwidth]{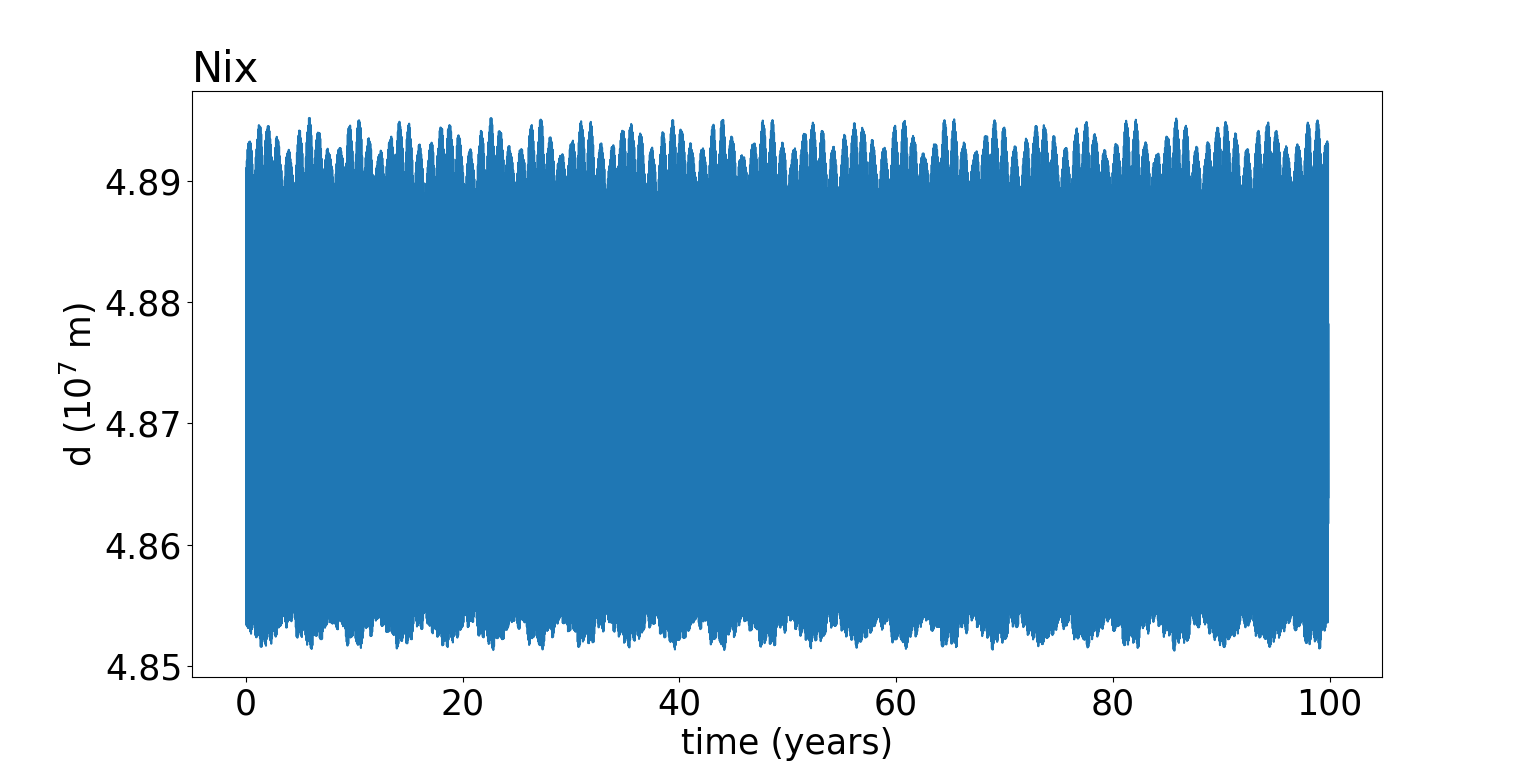} \\
        e\includegraphics[width=0.49\textwidth]{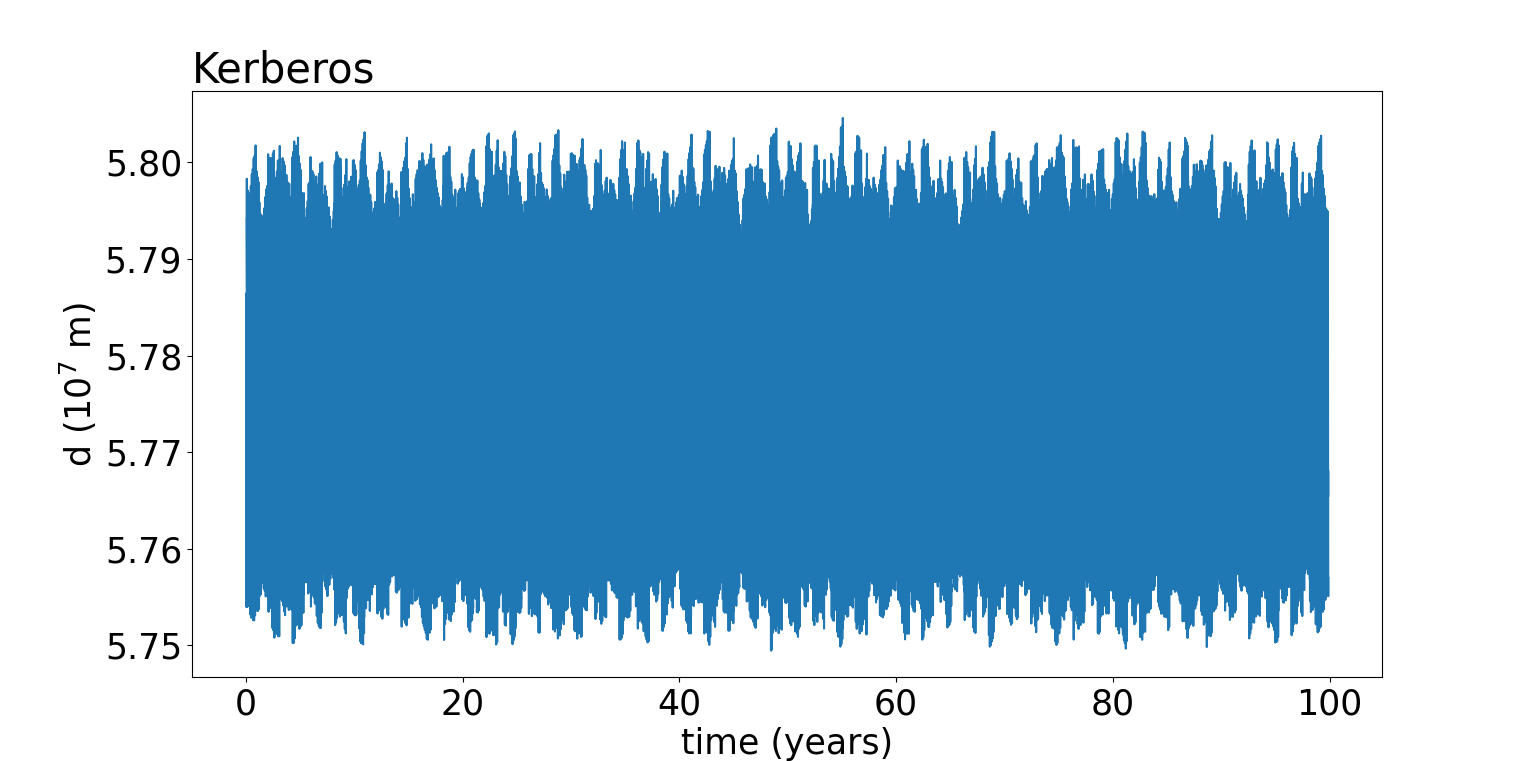}&     f\includegraphics[width=0.49\textwidth]{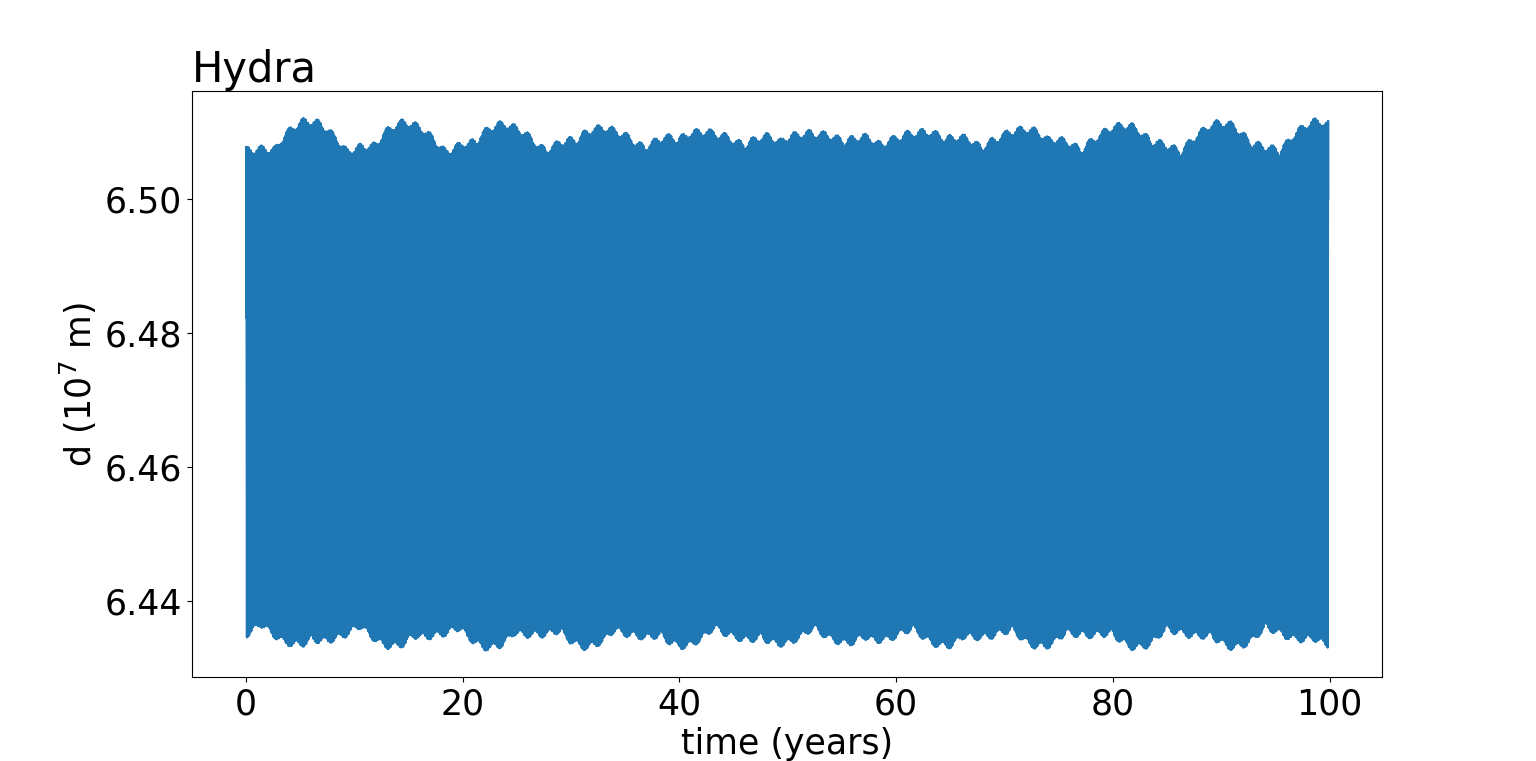}
        \end{tabular}
\caption{Variation of distance from barycenter ($d$) through a timespan of 100 years for Pluto, Charon, Styx, Nix, Kerberos and Hydra at respective panels a, b, c, d, e and f. The initial conditions employed correspond to case A.}
\label{fig:8}       
\end{figure*}

Furthermore, period values are not as distinct. There are discrepancies for the elements of other moons between the two papers as well, but Styx is the most prominent. Our suggestion is that it can be explained by the proximity of Styx to the center of mass; Styx is more intensely affected by the binary planet. This is the reason why it has the most irregular orbital pattern. Besides, Styx was detected for the first time 10 years ago (2012) and is the lightest body of the system, therefore it comes as no surprise that more uncertainties arise from its orbit. More observations over a longer timespan would give us a better assessment for the system. 

Lastly, we conclude that all moon orbits appear to be stable within uncertainties for at least several thousands years, without any indications for radical changes in their orbital behaviour (like collisions or close approaches deep inside mutual Hill radii, escapes to infinity or migrations). Determining the total lifetime of the system is beyond the scope of this work. However, large orbital deviations derive from our simulations in spite of adopting the most precisely calculated masses from photometry and distances from astrometry. Again, as we remark above, the time-changing Pluto's and Charon's potential primarily induces those discrepancies. In Figure \ref{fig:8}, the orbital distances for the case A are presented through a timespan of 100 years. Shorter- and larger-scale variations are visible, but are not large enough to indicate the crossing of any moon through the stability limits discussed in Section \ref{sec:3.2}. In spite of the observed amplitudes in the distance plot of each moon, Hydra remains well inside Szebehely's limit, since its maximum distance from the barycenter is still only $\sim 1\%$ of the critical distance. Likewise, Styx, at minimum orbital distance $4.207 \times 10^4$ km, stays beyond the innermost stable orbit ($3.915 \times 10^4$ km). Thus, at timescales like the ones we examine, the central binary cannot destabilize the system.

\section{Conclusions}
\label{sec:6}
In this paper, we have examined Pluto's and moons' motions and tried to model orbits around the binary planet. To this end, we have performed a series of n-body simulations and explored variations in positions and velocities through time. Since our aim was to study Pluto's moon system itself, our n-body integrator used only the five moons present in the vicinity of Pluto, thus neglecting the effects of the Sun or other heavy planets. As a result, we managed to inspect how much the central binary and other moons affect the orbit of one another. Furthermore, we have treated all objects as point masses, while they are non-spherical, thus a possible further improvement in this model could take this into account. 

We confirm that Keplerian orbital elements cannot be used to model circumbinary orbits. Published results from literature are averages for a long time of observations or simulations. We then assess that major uncertainties for an orbital characteristic do not only reveal observational limitations or simulation imperfection, but are in reality a result of the underlying specifications of the dynamical system; the non-axisymmetric potential of Pluto and Charon force moons into oscillations around their ``mean" orbits. It is therefore impossible to define an orbit around two masses that follows Kepler's laws. A key fact that points out the significant difference between an elliptic orbit around a single body and around a binary is that measured periods and semi-major axes are not in agreement. We speculate that this is not reflecting only the intrinsic challenges of studies, but it may well be related to the fact that the orbits, especially of the innermost moons, are well beyond ones appropriate to be characterized by osculating orbital elements.

Apart from the evident importance of this behaviour for solar system dynamics, we remark that such effects could be prominent in extrasolar planets orbiting binary stars. The fact that this system is stable suggests that from a dynamical perspective planets can survive a long time around binary stars. Yet, the variation of their barycentric distance due to the time-dependent, non-axisymmetric potential of the central binary makes places an additional challenge on determining the physical properties of such planets from their orbital parameters only. 

\begin{acknowledgements}
We thank an anonymous referee for a thorough and insightful review that has improved this paper. The numerical code used in this work was branched from an n-body code\footnote{ https://github.com/pmocz/nbody-python} created by Philip Mocz, that was based on the second-order leapfrog integration method (see, e.g. \cite{Springel:2021}). KNG acknowledges funding from grant FK 81641 "Theoretical and Computational Astrophysics", ELKE. The authors thank Apostolos Christou for insightful comments on this manuscript. Part of this paper was written on Mountain Aroania in Greece, within a short distance from the mythological dwelling of the deity Styx, and the Styx spring. 
\end{acknowledgements}



\bibliography{Bibtex.bib}   

\end{document}